\documentclass{article}

\usepackage{arxiv}

\usepackage[utf8]{inputenc} 
\pdfoutput=1
\usepackage[T1]{fontenc}    
\usepackage{hyperref}       
\usepackage{url}            
\usepackage{booktabs}       
\usepackage{amsfonts}       
\usepackage{amsmath}
\usepackage{nicefrac}       
\usepackage{microtype}      
\usepackage{graphicx}
\usepackage{doi}
\usepackage{bbold}
\usepackage[round]{natbib}
\title{Categorising the World into Local Climate Zones - Towards Quantifying Labelling Uncertainty for  Machine Learning Models}

\author{ Katharina Hechinger\\
	Department of Statistics\\
	Ludwig-Maximilians-University\\
	Munich, 80539 \\
	\texttt{katharina.hechinger@stat.uni-muenchen.de} \\
	\And
 Xiao Xiang Zhu \\
	Chair of Data Science in Earth Observation \\
	Technical University of Munich\\
	Munich, 80333 \\
	 \AND
  G\"{o}ran Kauermann \\
        Department of Statistics\\
        Ludwig-Maximilians-University\\
	Munich, 80539 
}

\date{}

\renewcommand{\headeright}{}
\renewcommand{\undertitle}{}
\renewcommand{\shorttitle}{}

\begin{document}
\maketitle

\begin{abstract}
Image classification is often prone to labelling uncertainty. To generate suitable training data, images are labelled according to evaluations of human experts. This can result in ambiguities, which will affect subsequent models. In this work, we aim to model the labelling uncertainty in the context of remote sensing and the classification of satellite images. We construct a multinomial mixture model given the evaluations of multiple experts. This is based on the assumption that there is no ambiguity of the image class, but apparently in the experts' opinion about it. The model parameters can be estimated by a stochastic EM algorithm. 
Analysing the estimates gives insights into sources of label uncertainty. Here, we focus on the general class ambiguity, the heterogeneity of experts, and the origin city of the images.
The results are relevant for all machine learning applications where image classification is pursued and labelling is subject to humans.
\end{abstract}



\section{Introduction}
\begin{figure}[h]
    \centering
    \includegraphics[width=\linewidth]{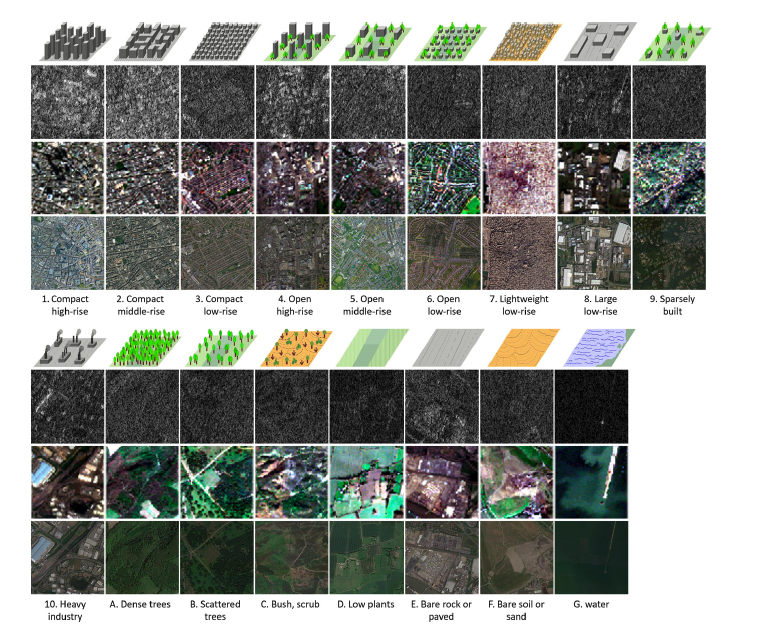}
    \caption[Exemplary image scenes of 17 LCZ classes.]{Example of image scenes of the 17 LCZ classes (upper image: Sentinel-1, middle image: Sentinel-2, lower image: high-resolution aerial image from Google).}
    \label{fig:LCZ_scheme}
\end{figure}

Machine Learning has achieved impressive standards in recent years. 
In particular, in image analysis and classification, deep learning has completely changed the way to approach image data. Today, machine learning is increasingly used for the classification of images, with applications for instance in medical image analysis, face recognition, machine vision and many more. In this paper, we focus on satellite images and their use to classify the world into so-called Local Climate Zones (LCZ) as a categorization of the surface.  The concept of LCZ, as proposed in \citet{Steward:2011}, has achieved a general standard in remote sensing and is based on the assumption that the structure of the landscape influences the local climate. The LCZ scheme categorizes the surface of the world into 17 classes that are supposed to influence local climate behaviour. The classes differ in surface structure (e.g. related to the density or height of trees and buildings) and surface cover (unsealed or sealed). A schematic description and exemplary satellite images are shown in Figure \ref{fig:LCZ_scheme}. This categorisation serves as an international standard for the mapping and analysis of urban areas and massive effort has been spent in developing algorithms that transform satellite images into an LCZ map. For this purpose, deep learning offers promising solutions to achieve high-quality maps and has already proven its utility in this regard, see e.g. \citet{Qiu:2019} or \citet{Qiu:2018}. \citet{Zhu:2022} combine earth observation data with deep learning and reveal detailed morphology of urban agglomerations across the globe. For an extensive overview of challenges, advances and resources of deep learning in the field of remote sensing, we refer to \citet{Zhu:2017}.

Machine learning is thereby based on labelled data, that is we are in the context of supervised learning, see e.g. \citet{Friedman:2001}. 
In this context, the problem of acquiring labels is very common and often solved by crowdsourcing, as introduced by \citet{Estelles:2012}. A lot of effort has already been spent in analysing the quality of such labels, e.g.\ by \citet{Raykar:2011} or \citet{Karger:2012} in the case of multi-class classification. \citet{Dawid:1978} also investigated the observed variation and its effect on the resulting measurements.
More recently, \citet{Chang:2017} developed a framework for improving the standard workflow of crowdsourcing by incorporating knowledge about labelling by experts. \citet{northcutt2021confident} also proposed Confident Learning in large (crowdsourced) databases with mislabelling, also referred to as ontological uncertainty. Though promising results can be achieved by exploiting the wisdom of the crowd, it is not suitable for our area of application. The classification of satellite images into local climate zones is non-trivial and relies on special knowledge and detailed classification instructions. Therefore, the insights from crowdsourcing theory are helpful to a certain degree but cannot be transferred directly for the classification of local climate zones.
In particular, our use case requires that experts label images by hand, classifying hundreds of images into one of the 17 categories. This process is apparently time-consuming and not without ambiguities. In fact, different experts come to different conclusions when classifying images. The quantification of remote sensing uncertainty is therefore particularly crucial as data sources are highly inhomogeneous and labelled image data are rare, see \citet{Russwurm:2020}. All in all, classifying satellite images into their corresponding LCZs demands a complicated and time-consuming annotation process. 

Using noisy or even deficient labels for the training of deep learning models leads to huge uncertainties and can result in serious challenges. Our problem concentrates on so-called label noise and we refer to \citet{frenay:2014} for an extensive survey. We are also faced with label ambiguity, where methods like label distribution learning have been introduced by e.g.\ \cite{Geng:2016}. Another approach is to incorporate the human component. \citet{Dgani:2018} discuss methods of training neural networks despite unreliable human annotations and \citet{Peterson:2019} incorporate this human uncertainty to increase the robustness of classification algorithms. \citet{Luo:2021} investigate label distribution learning also in the particular field of remote sensing.

The problem of labelling uncertainty goes well beyond the particular problem considered here. It is found also e.g. in medical image analysis as described in \citet{Zhang:2020} or \citet{Ju:2021}, face identification \citep{Kamar:2012} or more generally in crowdsourcing areas \citep{Phillips:2018}. In this work, we consider data, where each image has been classified by multiple experts
but the true class remains unknown. This relates to the setting of Latent Class models, as introduced by \citet{Lazarsfeld:1950}, where a set of observed variables is related to a set of latent variables. \citet{Goodman:1974} extended the original idea by using Maximum Likelihood methods and today, numerous variants of latent class analysis exist \citep{Magidson:2020}. These methods are helpful in many applications where the goal is to uncover hidden groups or structures in observed data.
In this work, we aim to quantify the uncertainty of the experts about some of the images by applying a classical finite mixture model. We refer to  \citet{McLachlanPeel:2000} or \citet{mclachlan2019finite} for a general description of the model class. See also 
\citet{fraley2002model} for the relation of mixture models and model-based clustering and \citet{cadez:2001} for an application to transaction data.
To link the application to mixture models we
assume a latent ground truth.
To be specific, we employ a multinomial mixture model and our ultimate goal is to estimate the "true" confusion matrix, i.e.\ without knowing the ground truth of an image. This will allow us to investigate the inevitable uncertainty in human image labelling. Moreover, we investigate if and how this uncertainty differs for images from different regions of the world, i.e.\ if and how the accuracy of annotation differs locally.

The quantification of uncertainty is receiving increasing interest in machine learning in recent years. We refer to \citet{Gawlikowski:2021} or \citet{Hullermeier:2021} for a general overview.
Typically, uncertainty is decomposed into two parts: {\sl aleatoric uncertainty}
and {\sl epistemic uncertainty}, sometimes also labelled as irreducible and reducible uncertainty. Such decompositions are not uniquely defined, and here we focus on an additional layer of uncertainty, which is often omitted, namely that the ground truth remains unknown. In our case, for each satellite image, we only have the annotations given by the human experts but the true LCZ is not given. 


The paper is organized as follows. In Section 2 we give a detailed description of the data at hand and describe the annotation process that has been applied to generate the data. In Section 3 we introduce our statistical approach and the models used to quantify the labelling uncertainty. Section 4 discusses the results of the particular data set at hand. Section 5 concludes the paper.

\section{Data}

We will analyse label uncertainty based on the earth observation benchmark data set So2Sat LCZ42 (\citealp{so2sat:2021}). For a detailed description of the full data set, we refer to \citet{Zhu:2019}. It comprises the LCZ labels of Sentinel-1 and Sentinel-2 image patches in 42 urban agglomerations across the globe. The images come in so-called patches, each covering an area of 320m by 320m. Figure \ref{fig:LCZ_scheme} shows an illustration of the LCZs, as well as examples of corresponding remote-sensing image patches.
The data set was created by a complicated and labour-intensive labelling project. For selected cities, polygons of different sizes were extracted, delineated such that the surface was largely homogeneous within each polygon. Within these polygons, equidistant images were then selected and initially labelled by a panel of two experts, which also
used auxiliary data 
such as high-resolution satellite images from Google Earth.
This procedure resulted in "clusters" of images, manually labelled by a larger panel of 11 experts. We here focus exclusively on these 11 votes per image. The produced labels do not serve as final labelling but rather as a validation stage. At this stage, one aims to assess overall labelling quality by comparing the opinion of the experts to the previously found label and possibly correcting it accordingly. A detailed layout of the labelling procedure and the validation stage can be found in \citet{Zhu:2019}.
Finding the suitable LCZ for a polygon is impossible for a layperson and even labour-intense and non-trivial for trained experts. As the definition of LCZs is very vague in its nature, a rigorous labelling workflow and decision rules had to be designed to ensure the highest labelling quality possible. 
Overall, we look at 159581 images from 9 cities leading to a data structure as sketched in Table \ref{tab:one_hot_example}.

\begin{table}
\caption{Sketch of the database.}
\label{tab:one_hot_example}
\flushleft
\fbox{%
\begin{tabular}{c|ccccccccccccccccc|c}
	Image ID & 1 & 2 & 3 & 4 & 5 & 6 & 7 & 8 & 9 & 10 & A & B & C & D & E & F & G & City
	\\ \hline
	1&0 & 0 & 0 & 0 & 0 & 0 & 0 & 0 & 0 & 0 & 0 & 0 & 0 & 0 & 0 & 0 & 11 & Berlin\\
	2&0 & 0 & 0 & 0 & 0 & 0 & 0 & 0 & 4 & 0 & 0 & 0 & 0 & 7 & 0 & 0 & 0 & Berlin \\
	... & \multicolumn{17}{c|}{...} & ...    \\
	3650 &0 & 0 & 0 & 0 & 0 & 0 & 0 & 0 & 0 & 0 & 0 & 0 & 0 & 6 & 0 & 0 & 5 & Zurich  \\
	... & \multicolumn{17}{c|}{...} & ... 
\end{tabular}}
\end{table}

\begin{figure}
    \centering
    \includegraphics[width=0.49\textwidth]{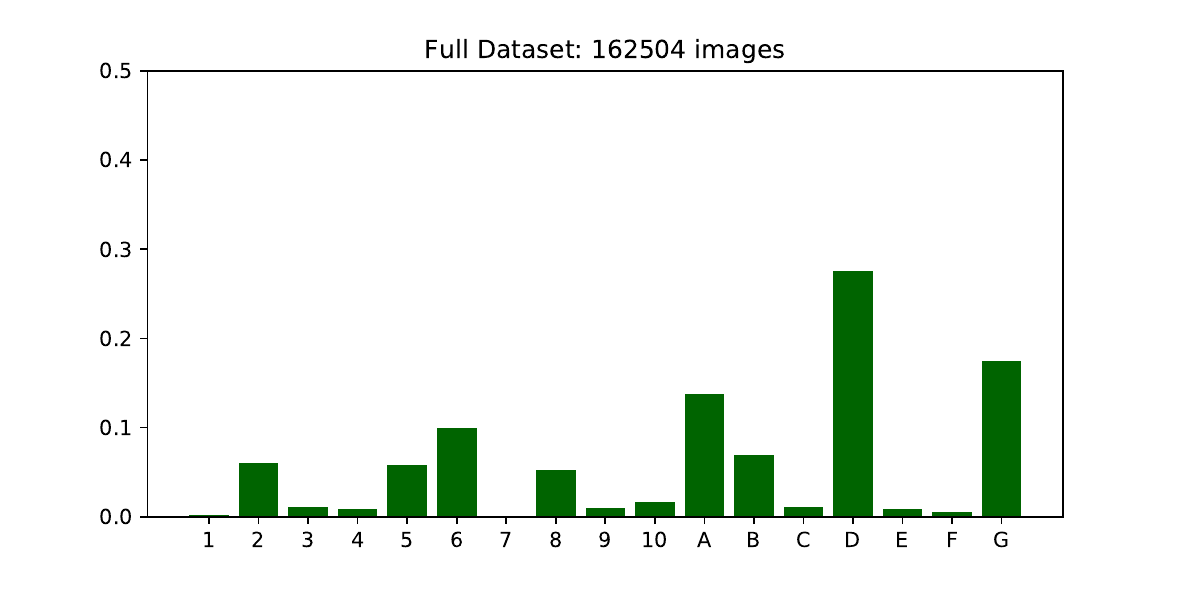}
    \includegraphics[width=0.49\textwidth]{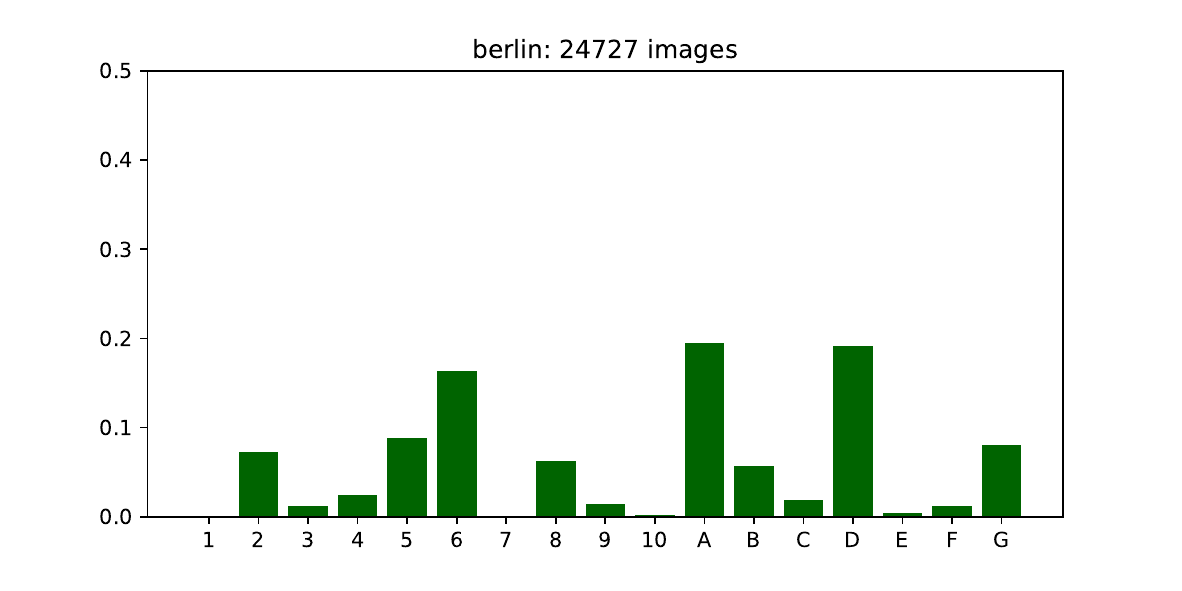} \\
    \includegraphics[width=0.49\textwidth]{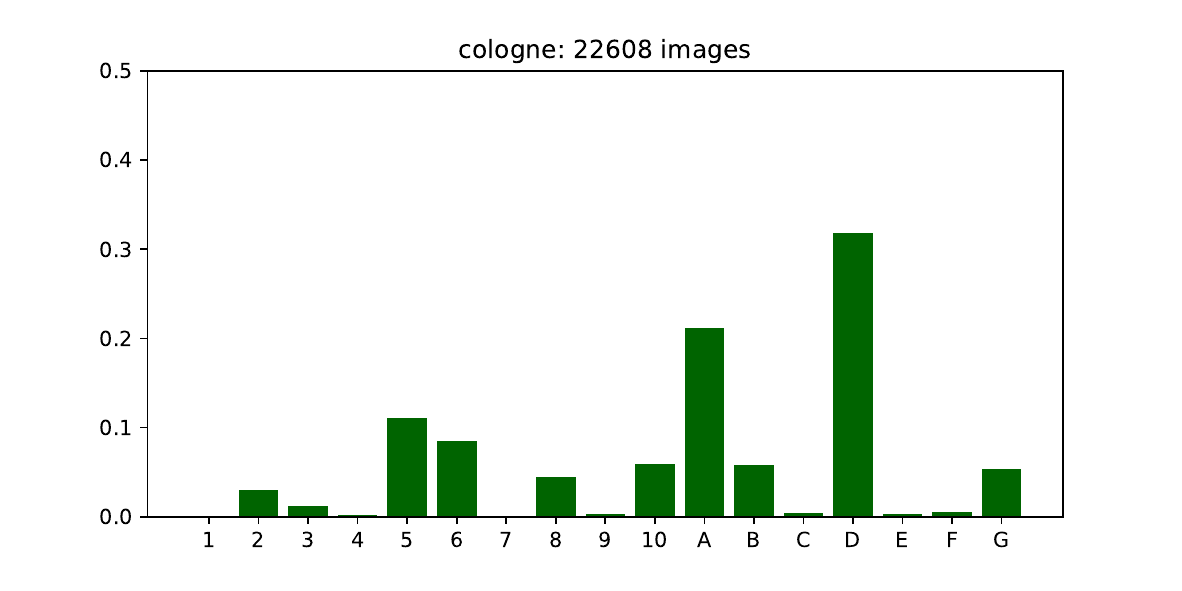}
    \includegraphics[width=0.49\textwidth]{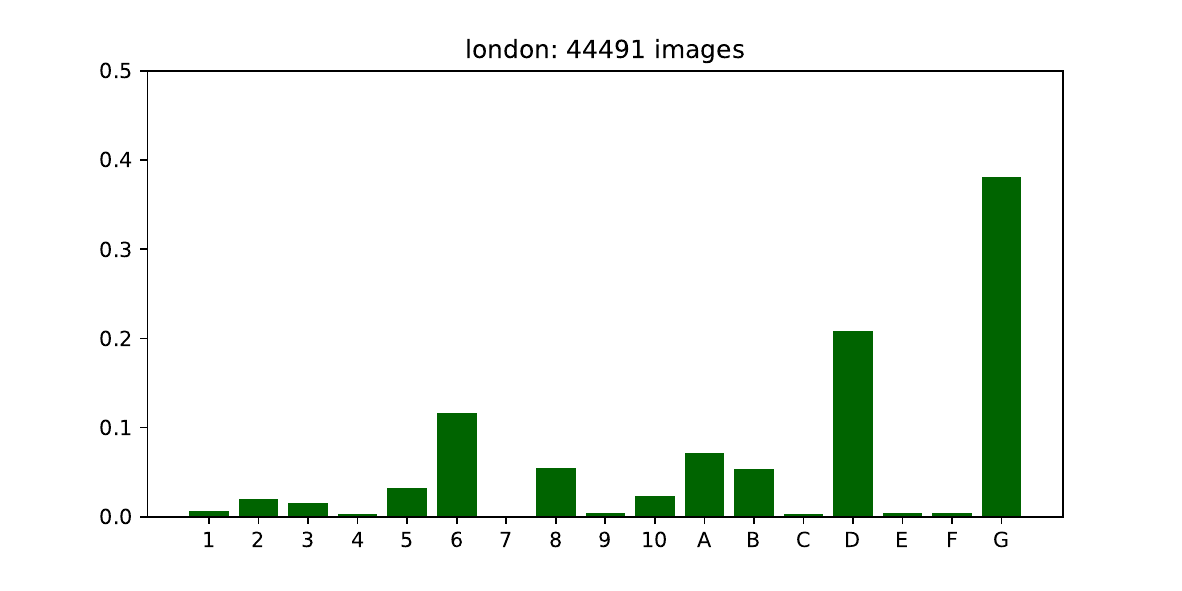} \\
    \includegraphics[width=0.49\textwidth]{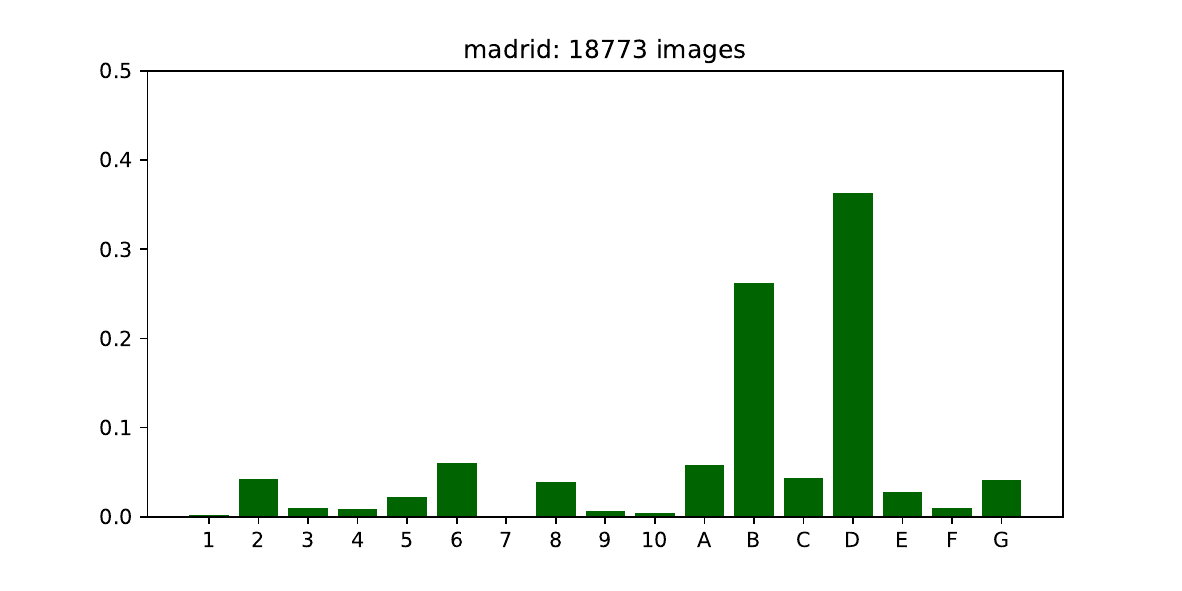}
    \includegraphics[width=0.49\textwidth]{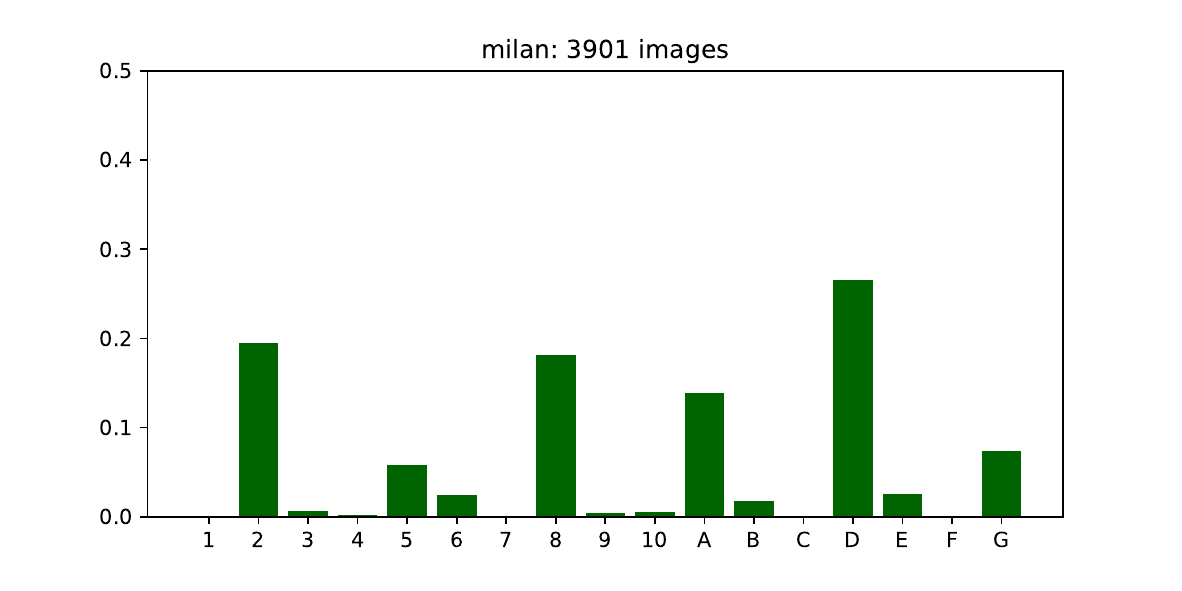} \\
    \includegraphics[width=0.49\textwidth]{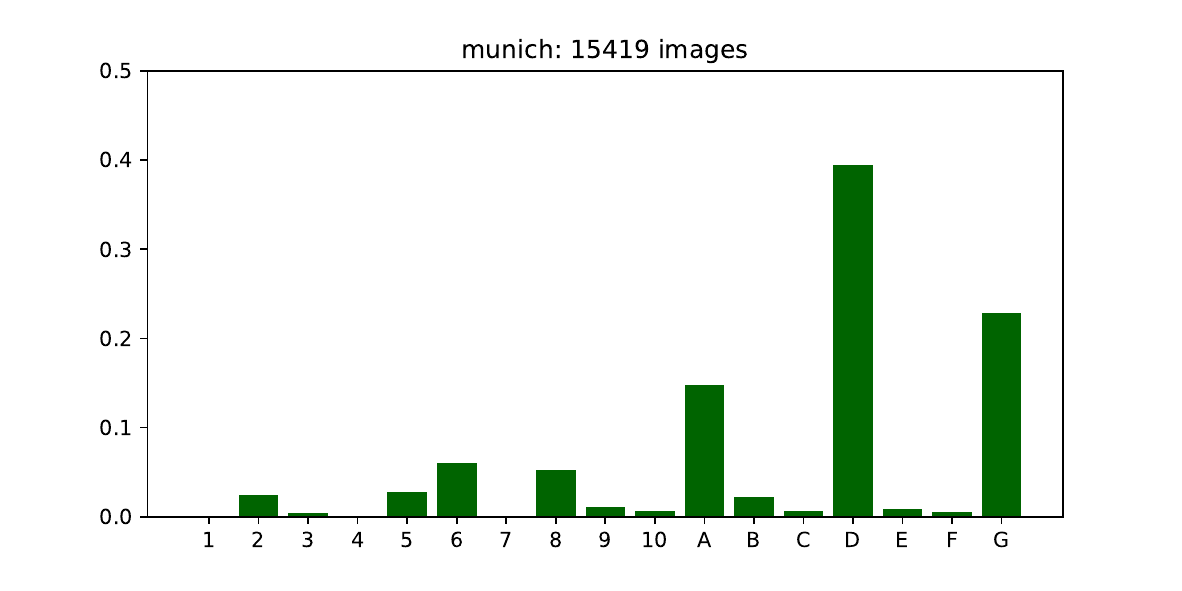}
    \includegraphics[width=0.49\textwidth]{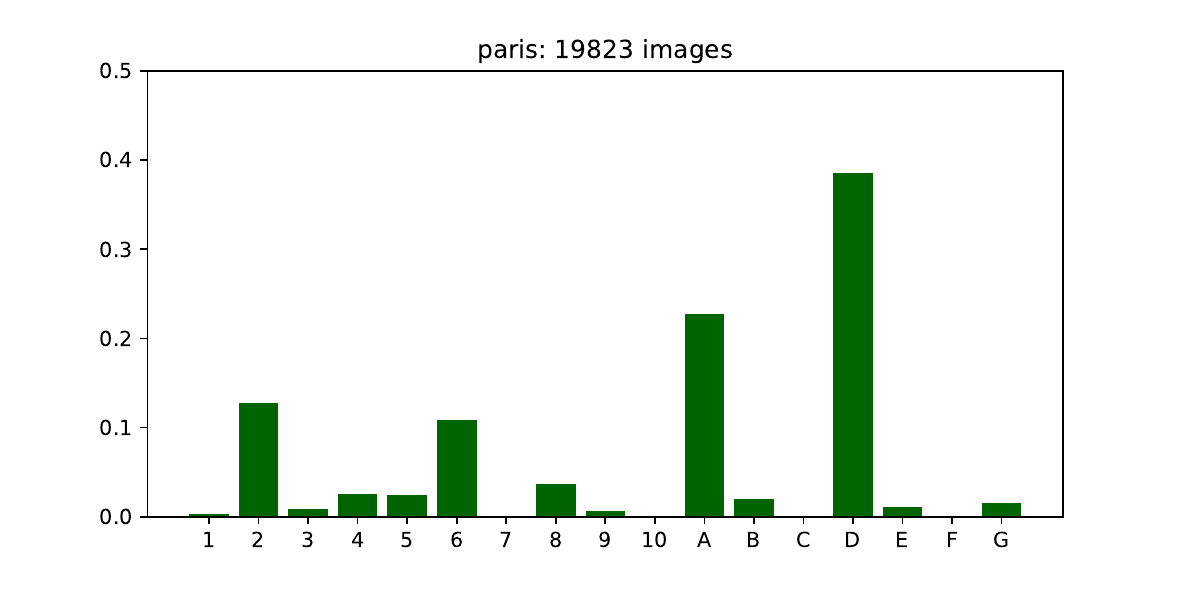} \\
    \includegraphics[width=0.49\textwidth]{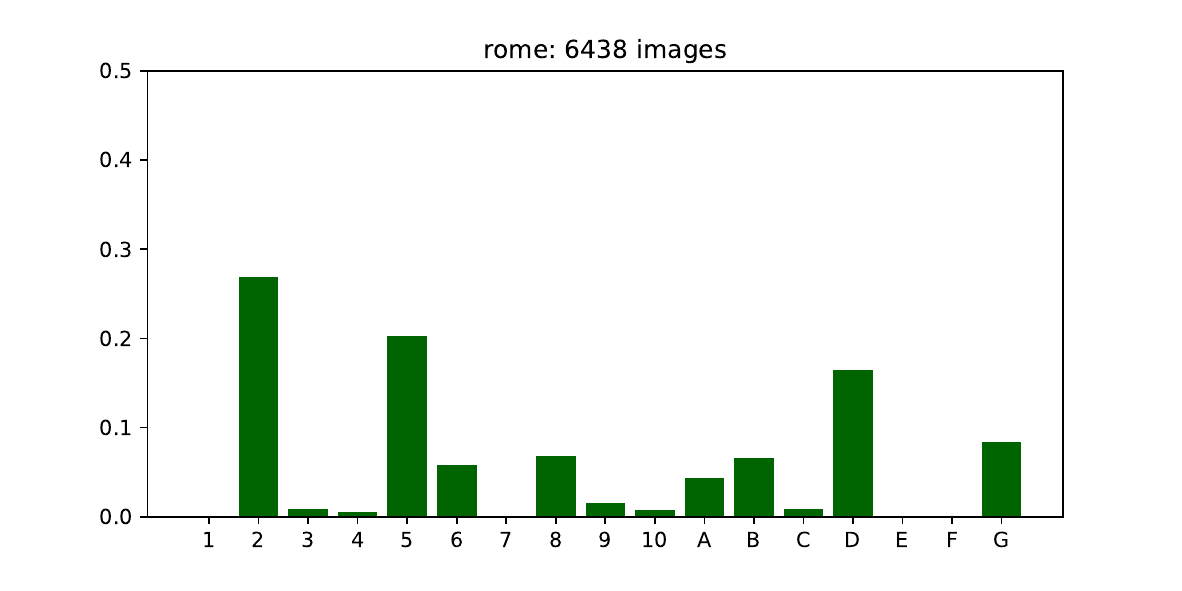}
    \includegraphics[width=0.49\textwidth]{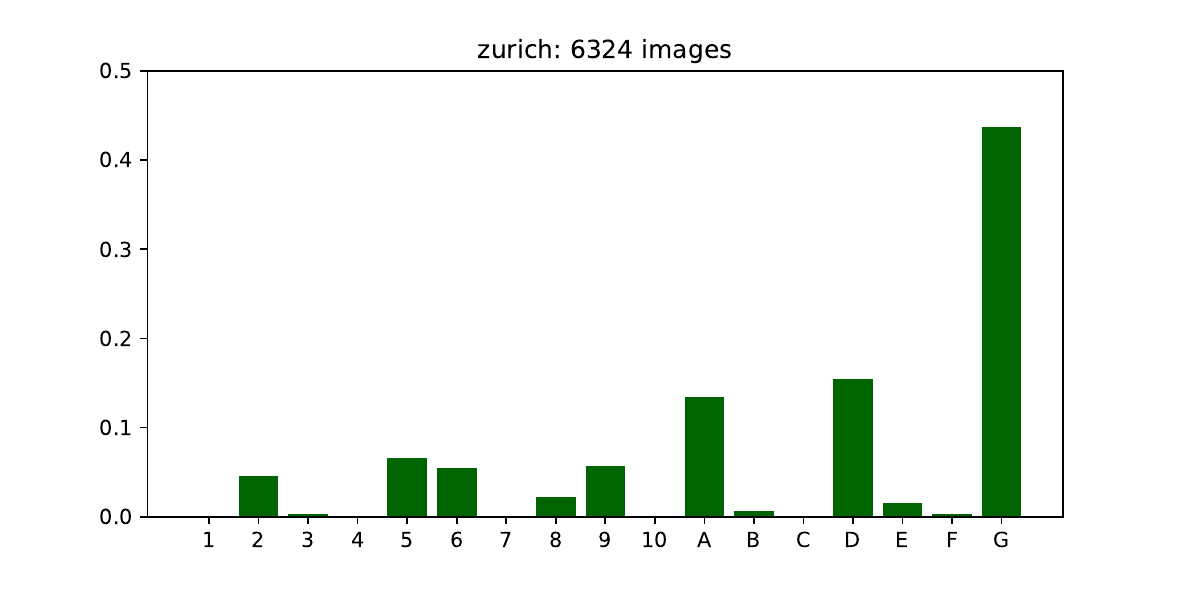}
    \caption[Class distribution of votes per city.]{Class distribution of the votes per city, along with the number of image patches per city. The top left figure shows the distribution for all images, the remaining figures show the distribution in Berlin (row 1), Cologne and London (row 2), Madrid and Milan (row 3), Munich and Paris (row 4), Rome and Zurich (row 5).  }
    \label{fig:City_class_distribution}
\end{figure}

The voting data already suggests some degree of certainty for the experts.
For 77.18 \% of the images, the experts agree on one single LCZ. We observe so-called 'voting patterns' for the other images. 11 experts voting for 17 classes leads to $11^{17}$ possible patterns, of which only 243 occur in the data set. This observation suggests that only a few classes are frequently confused, while others can always be distinguished. \\
The votes are not distributed evenly among the classes. As Figure \ref{fig:City_class_distribution} shows, the majority of votes were for classes A, D and G, whereas other classes hardly occur in the votings. 
Looking at the different cities, there is a large difference in not only the number of patches per city but also in the distribution of votes within the cities, see Figure \ref{fig:City_class_distribution}. We suspect a spatial correlation within the cities that influences the collection of votes. Looking at the plots, the distributions of votes in the different cities vary quite a lot. For example, class $G$ dominates in London or Zurich, while it hardly occurs in Paris. One should also note that the number of images inspected by the experts is also different in each city, which might impact the quality of the voting process.

\begin{figure}
    \centering
    \includegraphics[width=0.8\textwidth]{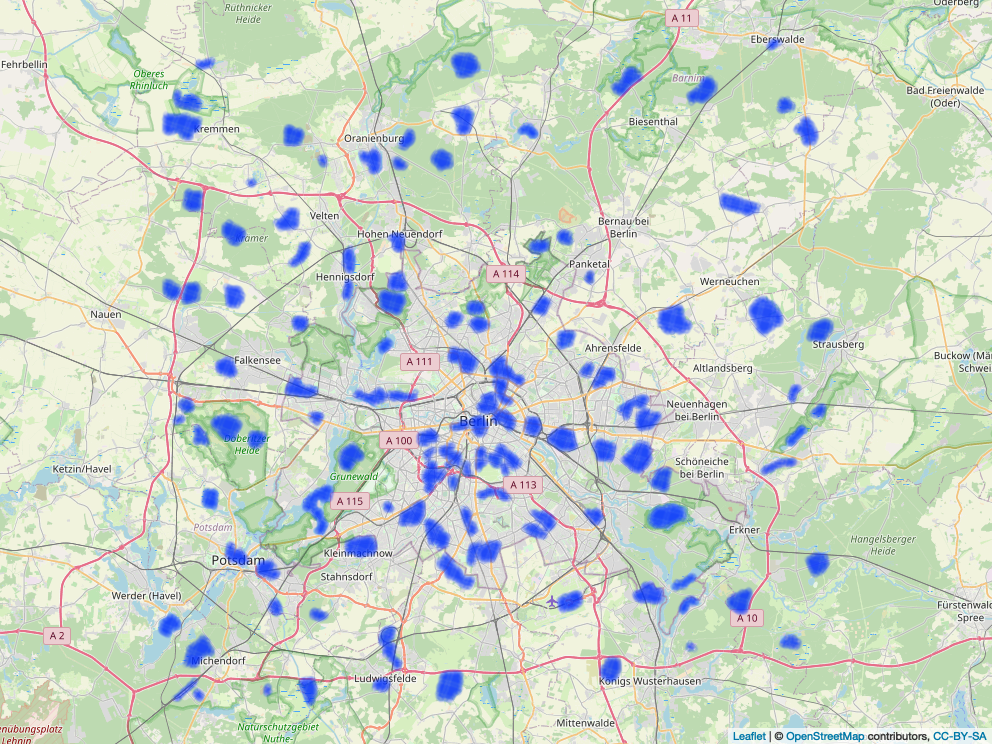}
    \caption[Spatial distribution of images and polygons across Berlin.]{ \label{fig:Berlin_geo} Spatial distribution of images and polygons across Berlin.}
\end{figure}

Another interesting aspect of the data is its clustered structure. The images were selected through cluster sampling of polygons including homogeneous areas. 
In Figure \ref{fig:Berlin_geo} we show exemplary the locations of the selected images in Berlin. The clustered structure is apparent. Other cities look comparable. 

\begin{figure}[h]
    \centering
    \includegraphics[width=0.8\textwidth]{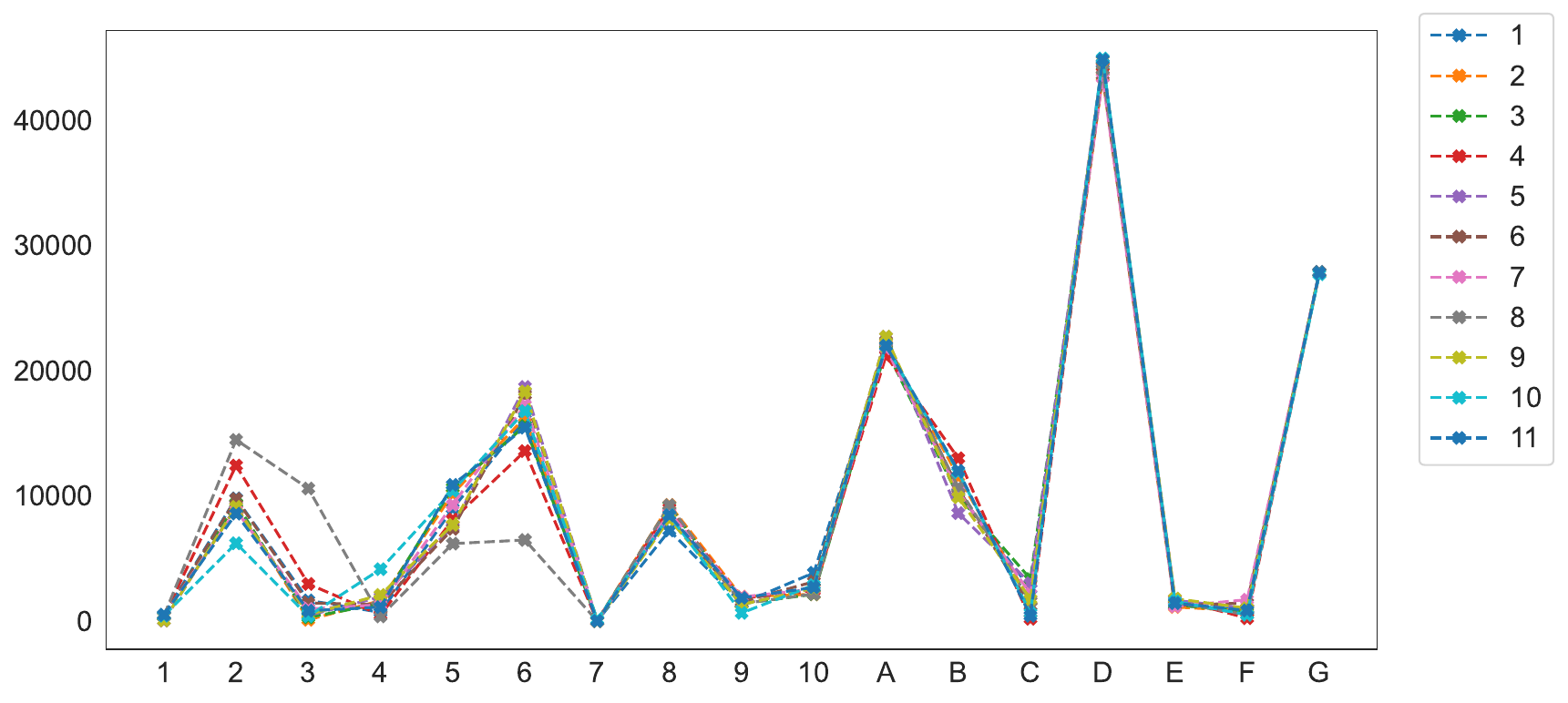}
    \caption[Distribution of votes by expert.]{Distribution of votes, split by expert.}
    \label{fig:class_distribution_experts}
\end{figure}

Finally, we look at the data from the voters' view. Figure \ref{fig:class_distribution_experts} shows a histogram of the votes cast by each expert. We recognise heterogeneity among the voters, where the voting behaviour differs mostly for urban classes (1-10), while the distributions for the non-urban classes A-G are pretty similar. We will therefore also aim to question if and how the voters' classification differs.

\section{Annotation Uncertainty}

\subsection{Description of Model}

To achieve the goal of exploring labelling uncertainty, we look at the votes cast by earth observation experts. Each image patch $i, i=1,...,n$ is assessed by a set of experts indexed with  $j, j=1,...,J$. The experts thereby classify each image individually into the LCZ $k$ where  $k=1,...,K$. The corresponding vote of the expert is denoted by $V_{j}^{(i)} \in \{1, \ldots , K\}$. It is notationally helpful to rewrite this vote into the $K$ dimensional indicator vector, which we denote in bold with  $\boldsymbol{V}_j^{(i)} = (\mathbb{1}\{V_{j}^{(i)} = 1\},..., \mathbb{1}\{V_{j}^{(i)} = K\})$,  with $\mathbb{1}\{\cdot\}$ as indicator function. This allows to accumulate the labellers' votes into the data points $\boldsymbol{Y}^{(i)} = (Y^{(i)}_1,...,Y^{(i)}_K)$ with $Y_k^{(i)} = \sum_{j=1}^J \mathbb{1}(V_j^{(i)} = k)$. This vector can be considered as the vote distribution for image $i$.

We assume further that each image comes from a single true class (=ground truth), which is a reasonable assumption based on the clustered data structure described above. Hence we assume that there is no ambiguity in the image class, but apparently, there are ambiguities in the voters' opinions about this class. We denote with $Z^{(i)} \in \{ 1, \ldots , K\}$ the true class of image $i$, which apparently remains unknown. Like above, we can reformulate the true class as $K$ dimensional index vector  $\boldsymbol{Z}^{(i)} =  (\mathbb{1}\{Z^{(i)} = 1\},..., \mathbb{1}\{Z^{(i)} = K\})$. Our intention is now to get information on $Z^{(i)}$ or $\boldsymbol{Z}^{(i)}$, respectively, given the voters' distribution $\boldsymbol{Y}^{(i)} $. We will therefore apply Bayesian reasoning using a Mixture Modell approach, which requires formulating a distribution framework. For the true classes, we assume a multinomial distribution, also called prior distribution, i.e. $$\boldsymbol{Z}^{(i)} \sim Multi(\boldsymbol{\pi}, 1) \mbox{ i.i.d.}, $$
where $\boldsymbol{\pi} = (\pi_1,...,\pi_K)$ with $\pi_k$ as so called prior probability that image $i$ is from the true LCZ $k$ for $k=1,\ldots , K$.  Given the true class of the image, we further assume that the labellers' votes also follow a multinomial distribution, i.e. 
\begin{align}
	\label{eq:model1}
	\boldsymbol{Y}^{(i)}|Z^{(i)} \sim Multi(\boldsymbol{\theta}_{Z^{(i)}}, J),
\end{align} 
where $\boldsymbol{\theta}_{Z^{(i)}} = (\theta_{Z^{(i)}1}, \ldots , \theta_{Z^{(i)}K})$. The parameters express the probabilities of voting for classes $1,...,K$ given the true class is $Z^{(i)} \in \{1,...,K\}$. We collect the coefficients into the matrix
$$
\Theta = ( \theta_{pk}, p,k = 1, \ldots K)
$$
which will be estimated from the data. Again, $\theta_{pk}$ refers to the voting probabilities, given $Z^{(i)} = p$.
We will also demonstrate how to estimate vector $\boldsymbol{\pi}$ if no prior knowledge about the general distribution of the LCZ is given or if any prior knowledge is intended to be ignored. Note that this approach corresponds to empirical Bayes estimation by estimating the prior to maximize the marginal likelihood, see e.g. \citet{robbins1992empirical}.

Given that the true classes are unobserved, we are in the framework of mixture models. We obtain the likelihood contribution of the $i$-th image by summing over all classes, that is 
\begin{align}
	\label{eq:lik1}
	P(\boldsymbol{Y}^{(i)}, \boldsymbol{\pi}, \Theta ) = \sum_{k=1}^K \pi_k P(\boldsymbol{Y}^{(i)},\boldsymbol{\theta}_k) 
\end{align}
where $\boldsymbol{\theta}_k$ is one column of the true confusion matrix and the probability in the sum results from the model (\ref{eq:model1}). Apparently, this is getting clumsy, so we apply the EM algorithm, or more precisely, a stochastic version of it.

\subsection{Stochastic EM Algorithm}
The main idea of the Expectation Maximisation (EM) algorithm, introduced by \cite{Dempster:1977}, is that the latent image class $Z^{(i)}$ is replaced by its expected value, given the data and the current estimates. This gives complete data so that the above estimates can be easily derived. These steps are carried out iteratively.  While the EM algorithm in general is a handy tool, it is also very slow and numerically intense. In fact, in our example, we would need to calculate the posterior expectation for over 200,000 images. Instead, we make use of the Stochastic EM algorithm (SEM) as proposed in \citet{Celeux:1996}. Here, the E-step is replaced by a simulation step, leading to simulated true image classes and hence allowing for simple estimation. Like the EM algorithm, one iterates between two steps to estimate the unknown parameters. \\

Let $\hat{\boldsymbol{\pi}}_{<t>}$ and $\hat{\Theta}_{<t>}$ be the estimates in the $t$-th iteration step of the algorithm. Taking these parameters we can calculate the posterior probabilities 
\begin{align*}
	\tau_{<t>l}^{(i)} &= P(Z^{(i)} = l|\boldsymbol{Y}^{(i)};
	\boldsymbol{\hat{\pi}}_{<t>}, \hat{\Theta}_{<t>}) = \frac{P(Z^{(i)}=l;\hat{\boldsymbol{\pi}}_{<t>})
	P(\boldsymbol{Y}^{(i)}|Z^{(i)}=l;\hat{\Theta}_{<t>})}{P(\boldsymbol{Y}^{(i)}; \boldsymbol{\hat{\pi}}_{<t>}, \hat{\Theta}_{<t>})} \\
	&= \frac{\hat{\pi}_{<t>l} P(\boldsymbol{Y}^{(i)};\hat{\boldsymbol{\theta}}_{<t>l})}{\sum_{l'=1}^K \hat{\pi}_{<t>l'} P(\boldsymbol{Y}^{(i)};\hat{\boldsymbol{\theta}}_{<t>l'})}.
\end{align*}

The simulation-based E-step is now carried out by drawing
\begin{align*}
	Z^{(i)}_{<t>} \sim Multi( \boldsymbol{\tau}^{(i)}_{<t>},1),
\end{align*}
where $\boldsymbol{\tau}^{(i)}_{<t>} = 
(\tau^{(i)}_{<t>1}, \ldots , \tau^{(i)}_{<t>K})$. We obtain complete data with these simulated true classes, leading to new estimates based on the complete likelihood.


Using this standard SEM procedure, we produce a chain of estimates (or simulated values) at each iteration, namely $ (\hat{\Theta}_{<t>})_{t \geq 0} \mbox{ and } (\hat{\boldsymbol{\pi}}_{<t>})_{t \geq 0}$ and therewith $(\hat{\boldsymbol{\tau}}_{<t>})_{t \geq 0}$. The final estimate can then be calculated as the mean value of the produced estimates starting at iteration $t_0$,  the end of the burn-in, i.e.\ the mean parameter resulting from the last $T-t_0$ iterations. For the parameter $\Theta$ describing the voting probabilities this results in 
$$ \hat{\Theta}_{final} = \frac{1}{T - t_0}\sum_{t=t_0}^T \hat{\Theta}_{<t>}. $$


The stochastic version of the EM has two advantages. First, it is numerically more straightforward, though it requires additional computation. Secondly, we can directly quantify the uncertainty of the estimates. We are interested in the estimation variance of the parameters, primarily of course in the estimate of the (mis)classification matrix $\Theta$. We refer to  Rubin's formula resulting from multiple imputations, see \citet{Rubin:1976} and \citet{Little:2002}. 
Note that the matrix estimate of $\Theta$ does not have full rank since the rows sum up to one. We therefore drop the last column and write the matrix estimate into a vector $\hat{\boldsymbol{\vartheta}} = (\hat{\boldsymbol{\theta}}_{1,-K}^T, \ldots, \hat{\boldsymbol{\theta}}_{K,-K}^T)$, where $\hat{\boldsymbol{\theta}}_{l,-K}$ is the $K-1$ dimensional subvector resulting from the first $K-1$ columns of  $\hat{\boldsymbol{\theta}}_l$. We obtain
\begin{align}
	\label{eq:rubin}
	Var(\hat{\boldsymbol{\vartheta}}) = E_Z( Var(\hat{\boldsymbol{\vartheta}}|Z)) 
	+ Var_Z ( E(\hat{\boldsymbol{\vartheta}}|Z))
\end{align}
where the subscript $Z$ refers to expectation and variance with respect to the latent classes $Z^{(i)}$ for $i=1, \dots, n$. Note that for given $Z$ we are in a complete data scenario and it is not difficult to show that in this case subvectors  $\hat{\boldsymbol{\theta}}_{l,-K} $ and $\hat{\boldsymbol{\theta}}_{l',-K} $ of $\hat{\boldsymbol{\vartheta}} $ are independent for $l \ne l'$. This leads to the variance
\begin{align*}
	Var(\hat{\boldsymbol{\theta}}_{l,-K}|Z) & = 
	\frac{ \mbox{diag}(\boldsymbol{\theta}_{l,-K}) - \boldsymbol{\theta}_{l,-K}^T \boldsymbol{\theta}_{l,-K}  }
	{\sum_{i=1}^n \mathbb{1}\{Z^{(i)} = l)},
\end{align*}
which is estimated in the $t$-th iteration step by replacing $\boldsymbol{\theta}_{l,-K} $ through its estimate, i.e.
\begin{align*}
	\widehat{Var}(\hat{\boldsymbol{\theta}}_{l,-K}|Z_{<t>}) =
	\frac{ \mbox{diag}(\boldsymbol{\hat{\theta}}_{<t>l,-K}) - \boldsymbol{\hat{\theta}}_{<t>l,-K}^T \boldsymbol{\hat{\theta}}_{<t>l,-K} }
	{\sum_{i=1}^n \mathbb{1}\{Z^{(i)}_{<t>} = l\}}.
\end{align*}

Replacing now the expectation in (\ref{eq:rubin}) through the simulated steps from the EM algorithm allows us to estimate the variance through
\begin{align*}
	\widehat{Var}(\hat{\boldsymbol{\vartheta}}) & =
	\frac{1}{T-t_0} \sum_{t=t_0+1}^{T}
	\mbox{blockdiag} 
	\left( \widehat{Var}(\hat{\boldsymbol{\theta}}_{l,-K}|Z_{<t>}) \right)
	 +   \frac{1}{T-t_0-1} \sum_{t=t_0+1}^{T}
	\left ( \hat{\boldsymbol{\vartheta}}_{<t>} - \hat{\boldsymbol{\vartheta}}_{final} \right) 
	\left ( \hat{\boldsymbol{\vartheta}}_{<t>} - \hat{\boldsymbol{\vartheta}}_{final} \right)^T 
\end{align*}
with obvious definition of $\hat{\boldsymbol{\vartheta}}_{final}$.

\subsection{Label Switching}
Like in every mixture model,  the resulting classes are subject to label switching, i.e.\ the numbering of the resulting classes does not match the original numbering of the LCZs. In other words, while the classes labelled by the voters have explicit meaning and therefore an interpretable order, the latent classes are subject to permutation and have no explicit interpretation.  For the mixture model, we assume 17 true classes which are ordered at convergence as  $C=\{C_1,C_2,...\}.$ On the observation side the experts categorise the satellite images into 17 classes denoted by  $L=\{L_1,L_2,...\}.$ We now need to match the latent classes $C$ to the labelled classes $L$. It is important to note that the labels of the clusters returned by the algorithm are unidentifiable.
Therefore, to ensure a clear assignment, we need a bijective function going from the cluster labels $C$ to the voter labels $L$. Or putting it differently, we need to construct a permutation $\sigma()$ on the numbers $\{1, \ldots K\} $ such that $\sigma(C_l) = k $ means that the latent class $C_l$ corresponds to the LCZ $L_k$. This could be achieved by looking at the posterior probability of the latent classes given the voters' opinions. Note that for a single vote $V^{(i)}$ we obtain
$
P(Z^{(i)} = l | V^{(i)} = k ) \propto
P(  V^{(i)} = k | Z^{(i)} = l ) \pi_l
$ or written in matrix form
\begin{align*}
	P(Z^{(i)} = \{1, \ldots , K\}^T | V^{(i)} = \{1 , \ldots , K\} ) &\propto \mbox{diag}(\boldsymbol{\pi}) \Theta^T 
\end{align*}
This suggests constructing the permutation $\sigma()$ such that its inverse fulfills
\begin{align}
\label{eq:permut}
\sigma^{-1}(k) = \mbox{arg max}_{l}\big(\mbox{diag}(\boldsymbol{\pi}) \Theta^T)\big). 
\end{align}
This still might not lead to a unique definition. We, therefore, apply rule (\ref{eq:permut}) in descending order of the relative frequency of the labellers' votes and choose the arg max from the not-allocated classes only. A detailed layout of the algorithm is provided in the Appendix. After finding the correct permutation of the numbers, we rename the original clusters according to the respective local climate zone label. 
\section{Sources of uncertainty in the votes}   

We are now in the position to approach different questions related to human annotation of satellite images. These are:
\begin{enumerate}
    \item How distinguishable are the LCZs in general, that is can we estimate the "true" confusion matrix?
    \item Is there an expert bias, that is are the experts heterogeneous or homogeneous with respect to the labelling?
    \item Is the voting behaviour influenced by geographic differences, that is does the "true" confusion matrix differ in the different cities where the data come from?
\end{enumerate}
All three questions are tackled subsequently.

\subsection{Ambiguity of LCZs}
A very general aspect for quantifying the uncertainty in the voting data set are the LCZs themselves. By looking at the definition and characterisation of the classes, it is obvious that some are very similar and might not be easily distinguishable, even for experts. This is for example the case for classes 3 and 7, which both describe urban low-rise environments.  Contrarily, there are LCZs that are easy to discover on images and that are likely to be never confused by humans, e.g.\ class 17 covering water areas. In general, it is presumably more difficult to distinguish urban classes, i.e.\ LCZs 1 to 10 than non-urban classes which are LCZs A to G, details can be found in the Appendix. \\
The parameter of main interest is the confusion matrix $\hat{\Theta}$, i.e.\ the estimated true confusion matrix of the classes. \\

To obtain a stable estimation and interpretable results, we restricted the estimation procedure described in the previous section to $K=16$ instead of $K=17$, omitting LCZ 7 (lightweight low-rise building types). This is reasonable, not only due to the semantic interpretation of "slums", which are very unlikely to occur at all in European cities but also necessary due to the lacking data basis. As mixture models are generally able to handle any arbitrary number of classes, including a class without sufficient observations or votes in this case, this will lead to instability and confusion in the estimation results. \\

\begin{figure}
\centering
    \includegraphics[width=\textwidth]{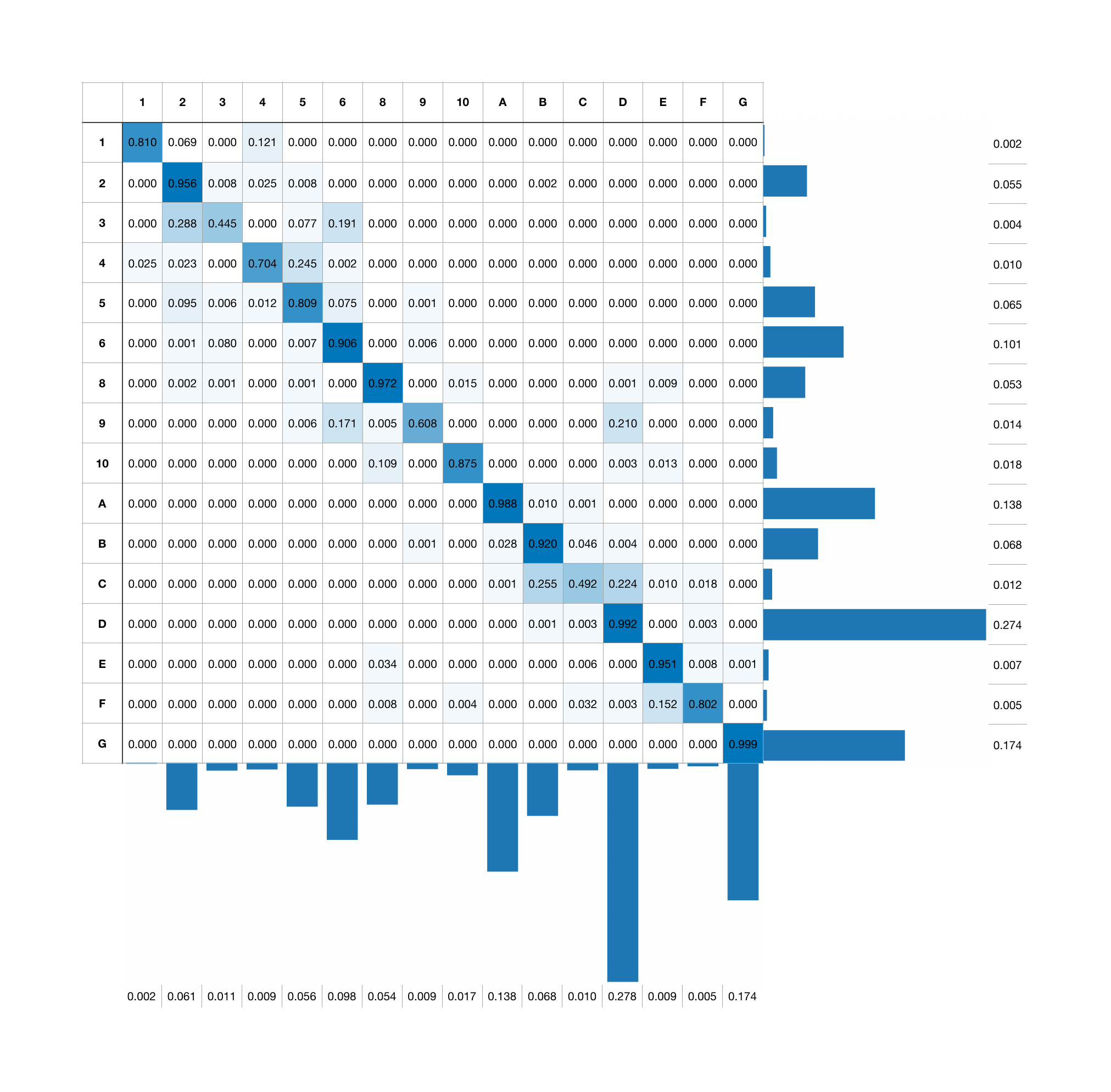}
    \caption[Estimated true confusion matrix.]{The matrix shows the true confusion of the voted local climate zones (columns) with the true classes (rows), along with the estimated prior probabilities on the right side and the relative vote frequency on the bottom of the plot.}
    \label{fig:true_confusion_full}
\end{figure}

Figure \ref{fig:true_confusion_full} shows the resulting estimate based on the full data. The entries on the diagonal contain the probability of correctly classifying images. In contrast, entries $\boldsymbol{\theta}_{lk}$ describe the probability of classifying an image that truly belongs to class $l$ into LCZ $k$ instead. Looking at the diagonal of the matrix, it is obvious that correct classification is highly dependent on the number of votes a class received from the experts. Apparently, Classes 2, 6, 8, 10, A, B, D, E and G are very well separable, whereas classes 3, 4, 9 and C are often not detected correctly. Note that said classes received a very small number of votes in the labelling process so the misclassification should not be over-interpreted. This can be seen from the frequency distribution indicated at the bottom of the plot and also the estimated prior distributions (right-hand side of the matrix), which show a strong tendency of the voters for classes A, D and G. Apart from correct classification probabilities, we can also detect classes that are not that easy to distinguish, like e.g. class 3, where the voting probabilities are distributed among classes 2, 3 and 6.

Generally, our results depend on the input votes as the algorithm can only detect classes where the data basis is sufficient. Furthermore, it should be mentioned here our true confusion matrix  is subject to the implemented label-switching process. As the multinomial mixture model produces "meaningless" clusters that must be assigned to LCZs afterwards, the resulting estimates and their interpretation are based on the assignment strategy, which might not be unambiguous. Generally, however, we obtain interpretable insight into the inevitable ambiguity when classifying LCZs. \\

\subsection{Expert Heterogeneity}

 As explained in the beginning, the task of classifying is not trivial, even for trained earth observation experts. Therefore, it is obvious that the human assessment causes some confusion and uncertainty within the data. The experts are assumed to be subject to some bias, that might impact or even skew the results. The model described in the previous section allows for assessing the impact of each individual expert and their heterogeneity. 
 As shown in Chapter 2, the distribution of votes for each expert can differ quite a lot, in particular for urban classes. Therefore it is worthwhile to investigate the individual impact of the voters further. If experts were homogeneous, their voting behaviour would not differ, and dropping the votes of one expert at a time should not change the final estimated distribution. \\
 Looking at Figure \ref{fig:class_distribution_experts}, we already get a general overview of the observed voting behaviour of the experts. While the distribution of votes is similar for all experts for the non-urban LCZs (A-G), the distribution varies noticeably for the urban classes (1-10). Therefore, it is worth further examining the voting behaviours and their impact on the results. \\
 The parameter of interest here is $\hat{\boldsymbol{\tau}}^{(i)}_{l}$, expressing the posterior probability of image $i$ to belong to the true class $l$ according to the applied model and algorithm. 
 In order to analyse the difference between the voting behaviours, we calculate the posterior probabilities excluding a single expert. This leads to 11  estimates $\hat{\boldsymbol{\tau}}^{(i)}_{(-j)l}$, where the bracketed index $-j$ refers to the excluded voter $j$. \\

\begin{figure}[h]
\centering
    \includegraphics[width = 0.49\textwidth]{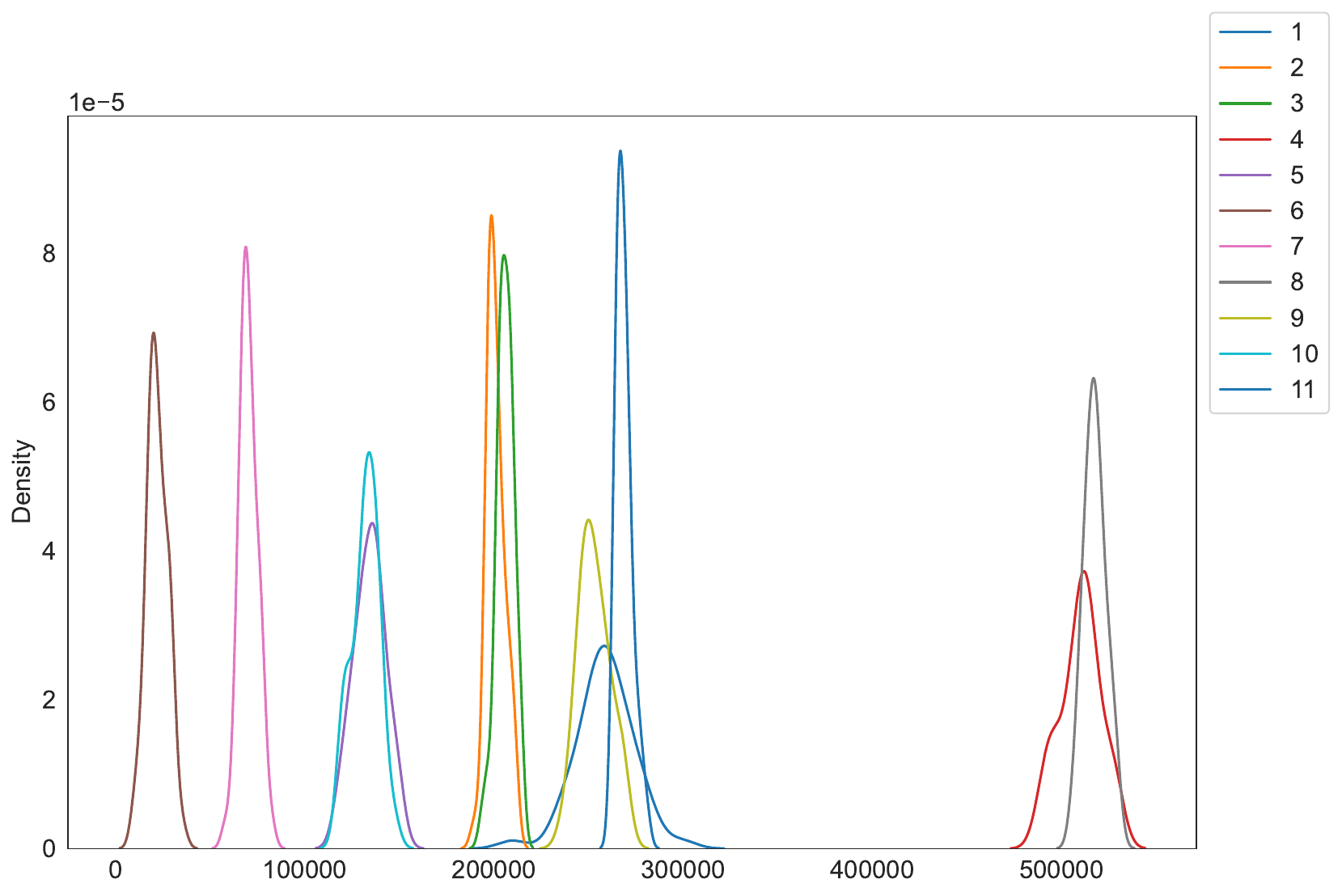}
    \includegraphics[width = 0.49\textwidth]{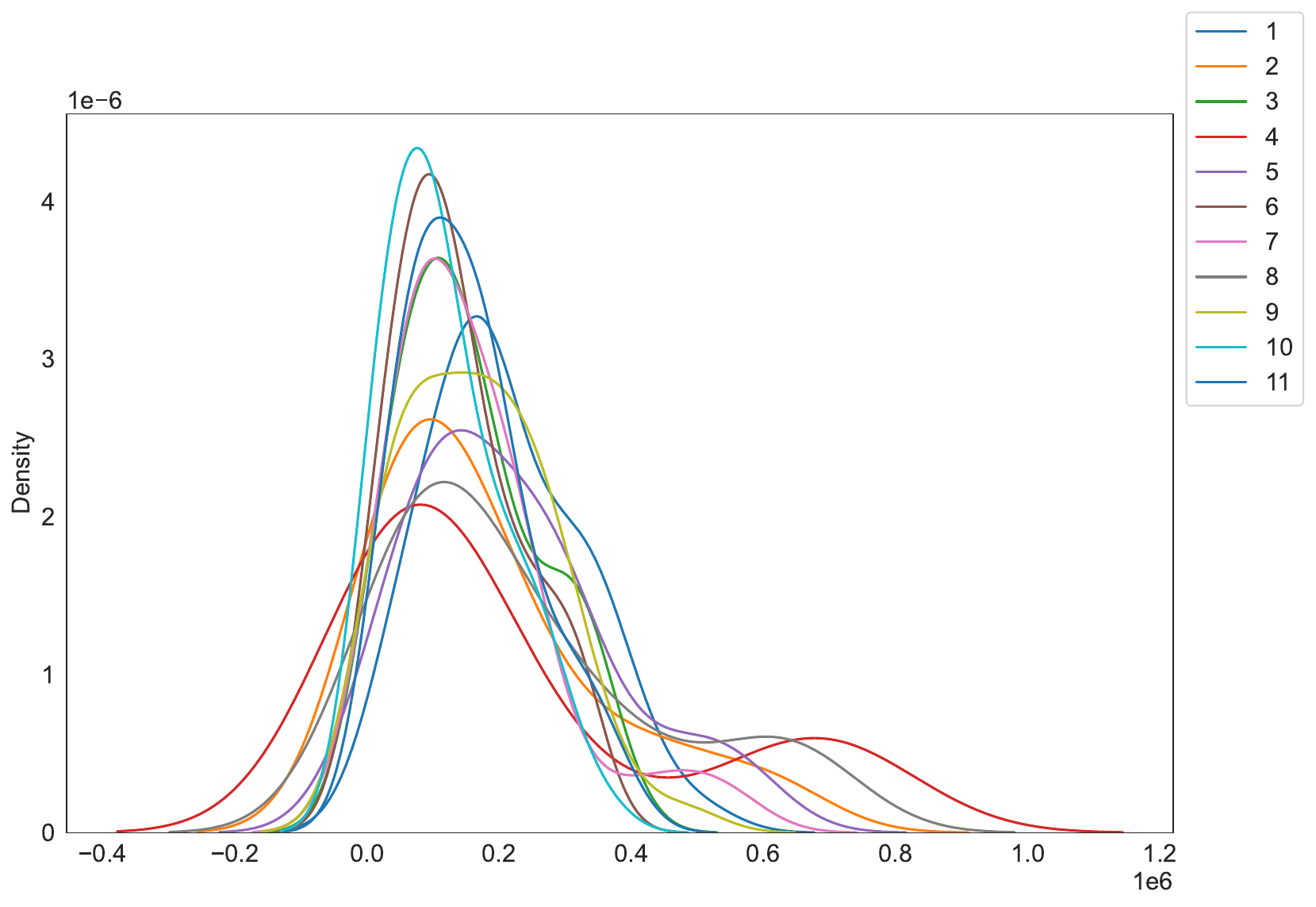}
    \caption[Densities of the goodness of fit statistics per expert.]{On the left-hand side, the densities of the differences of the bootstrapped goodness of fit statistics for each expert are shown. On the right-hand side, the densities belong to randomized versions, showing a clear difference between the actually observed and random votes.}
   \label{fig:expert_boot_density}
\end{figure}

A very straightforward way of quantifying heterogeneity is calculating the log-likelihood. Here, we assume that the distribution of the categorical variable is the same for all subgroups. In this particular application, the subgroups consist of 11 experts, which classify images into 16 LCZs. Assuming a multinomial distribution the $j$-th expert contributes to the negative log-likelihood through
\begin{align} 
\Lambda_j = -\sum_{i=1}^N \sum_{k=1}^K 1\{V_j^{(i)} = k\} \cdot \log(\tau_{k}^{(i)}),
\end{align}
where $0\cdot \log(0)$ is defined as 0. We replace $\tau_{k}^{(i)}$ by the estimate excluding the $j$ labeller and define the resulting negative log-likelihood as 
\begin{align} 
\hat{\Lambda}_j = -\sum_{i=1}^N \sum_{k=1}^K 1\{V_j^{(i)} = k\} \cdot \log(\hat{\tau}_{(-j)k}^{(i)}).
\label{eq:chi2}
\end{align}
Statistic $\hat{\Lambda}_j$ is a random quantity, which we will now explore through bootstrapping. We, therefore, draw $N$ images with replacement and denote with $V_j^{(i*)}$ the vote of the $j$-th labeller for the bootstrapped image $i*$. This leads to the bootstrap quantity 
\begin{align} \hat{\Lambda}^*_j = -\sum_{i=1}^N \sum_{k=1}^K 1\{V_j^{(i*)} = k\} \cdot \log(\tau_{(-j)}^{(i*)}).
\label{eq:chi2_bootstrap}
\end{align}
We repeat this step $B$ times to obtain $\hat{\Lambda}_j^{*b}$ for $b=1,...,B$. To put the magnitude of these bootstrapped values into perspective, we look at the absolute differences to the overall mean and define
\begin{align*} D_j := |\hat{\Lambda}_j - \frac{1}{J-1} \sum_{l=1, l\neq j}^J \hat{\Lambda}_l|
\end{align*}
and the resulting bootstrapped versions
\begin{align*} D_j^{*b} := |\hat{\Lambda}_j^{*b} - \frac{1}{J-1} \sum_{l=1, l\neq j}^J \hat{\Lambda}_l^{*b}|.
\end{align*}
If the experts are homogeneous, then $D_j$ should be in the range of 0, while if experts are heterogeneous, the range of $D_j$ will vary. We, therefore, look at the densities of the bootstrapped distances as shown on the left side of Figure \ref{fig:expert_boot_density}. The plots show a clear difference between the bootstrapped differences and therefore a clear separation of the experts regarding voting behaviour. To confirm this we run an additional bootstrap under the assumption that experts are homogeneous. To do so we conduct a randomized version of the former analysis by replacing the votes of expert $j$ in (\ref{eq:chi2_bootstrap}) with a random vote. In other words, for each image we randomly permute the experts. This leads to the densities displayed on the right side of \ref{fig:expert_boot_density}, showing homogeneous curves for randomized votes. \\

\begin{figure}[h]
\centering
    \includegraphics[width = \textwidth]{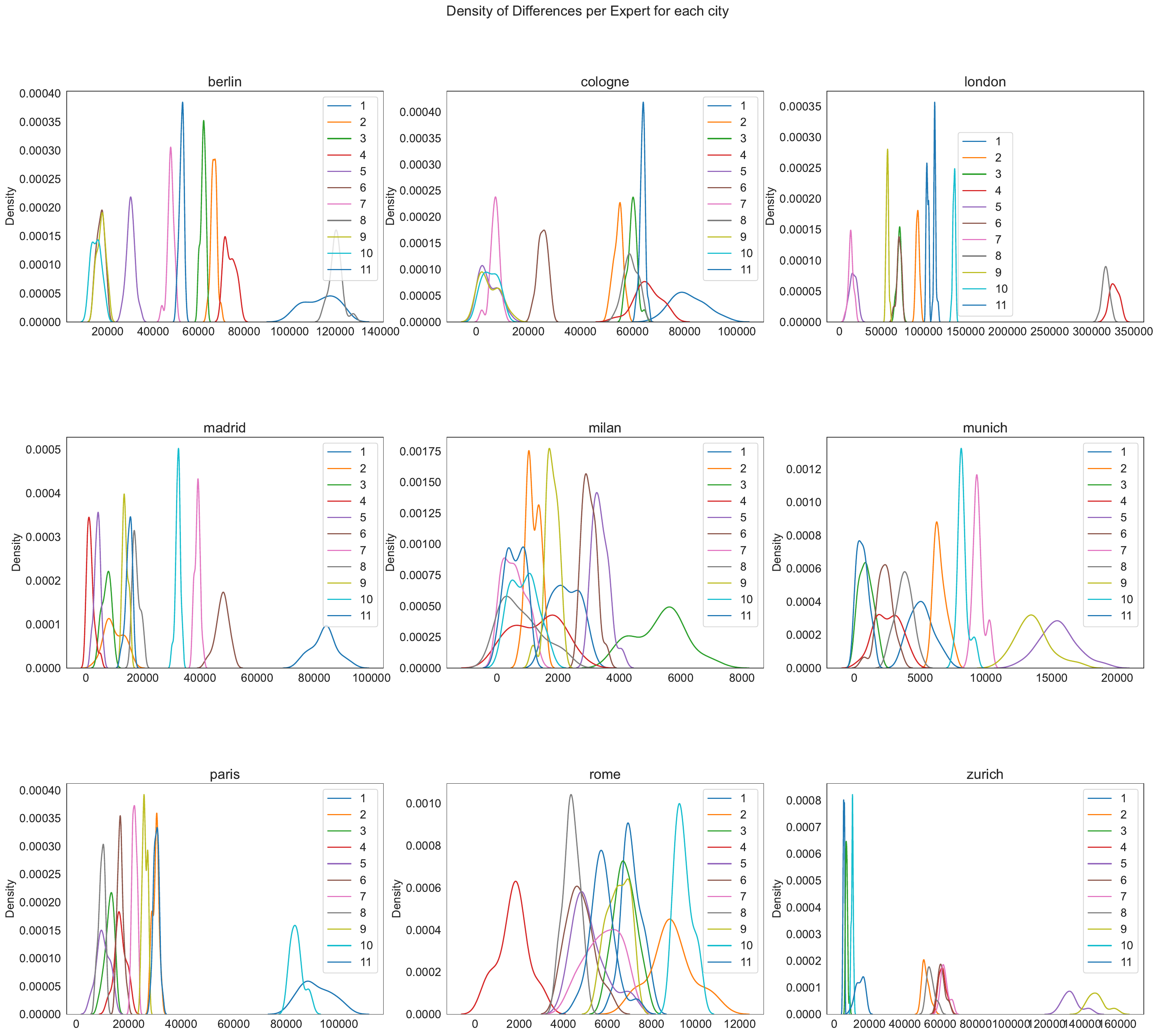}
    \caption[Densities of the differences of the bootstrapped goodness of fit statistics per expert per city.]{The plot shows the densities of the differences of the bootstrapped goodness of fit statistics for each expert, split up per city.}
   \label{fig:expert_boot_density_cities}
\end{figure}

We can investigate this further and check whether the experts' voting behaviour and their homogeneity differ for different cities. In Figure \ref{fig:expert_boot_density_cities} we show the same bootstrapped densities as described above for each city separately. It appears that the experts conduct their voting differently in comparison to the rest and this varies over the different cities. For some cities, the voting behaviour seems more homogeneous, which might however also mirror smaller sample sizes as in cities like Milan, Rome or Zurich. Overall, the analysis shows that while some of the experts align well with the overall voting behaviour, others conduct a more heterogeneous and individual voting, leading to confusion and an ambiguous majority voting. \\

Depending on the application at hand, expert heterogeneity is often not only accepted but also desired. In this particular case, experts received the same training on how to conduct classification of satellite images and are generally assumed to produce homogeneous votings.



\subsection{Geographic Differences}
The last question we want to consider is geographic variation, as an external influencing factor. The polygons used for the voting procedure come from 9 different European cities, which are known to be quite diverse in terms of structure and architecture. While this might be intended to cover all LCZs as well as possible, it complicates the assessment of the images. The question is whether earth observation experts have difficulties in assigning certain images to certain climate zones, depending on the respective region. Looking at Figure \ref{fig:City_class_distribution}, we see differences in the sample sizes and also the vote distributions in different cities. Additionally, Figure \ref{fig:expert_boot_density_cities} shows that the voting behaviour of experts is subject to the location of the images.
This brings us to the question of whether certain LCZs are harder to identify in some cities than in others. Apparently, this might have a relevant impact on the assessment by the experts and influence the voting behaviour, which leads to uncertainty in the training data set. However, one has to note here that the voting distribution always depends on the initial draw of images or polygons in each city. The pursued strategy might lead to imbalanced labels and therefore a bias in the voting probabilities. We here are however interested in the confusion matrix and whether this matrix differs in the different cities. Maps with the location of the used images are shown in the Appendix. \\

\begin{figure}[h]
    \centering
    \includegraphics[width = 0.494\textwidth]{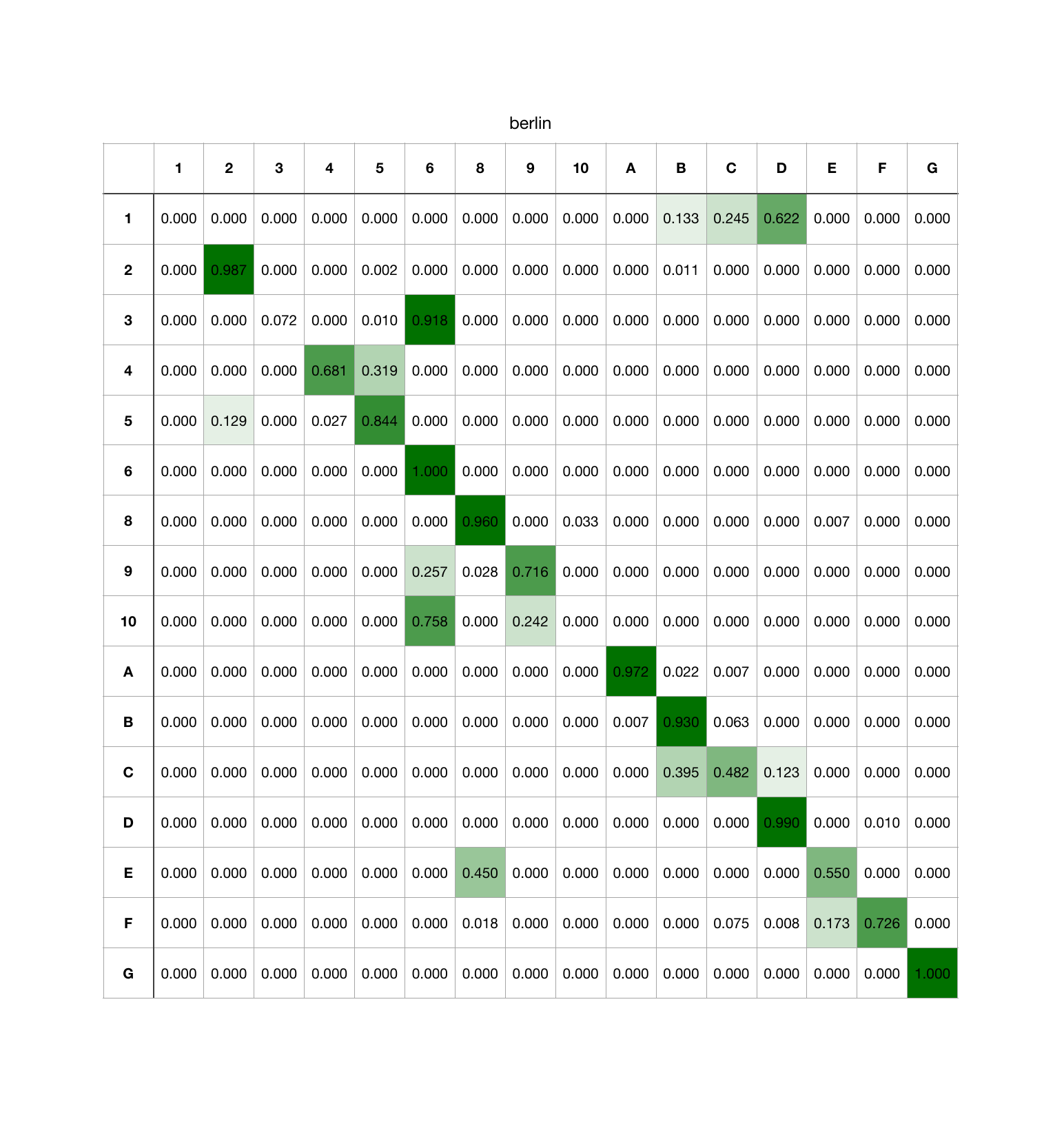}
    \includegraphics[width = 0.494\textwidth]{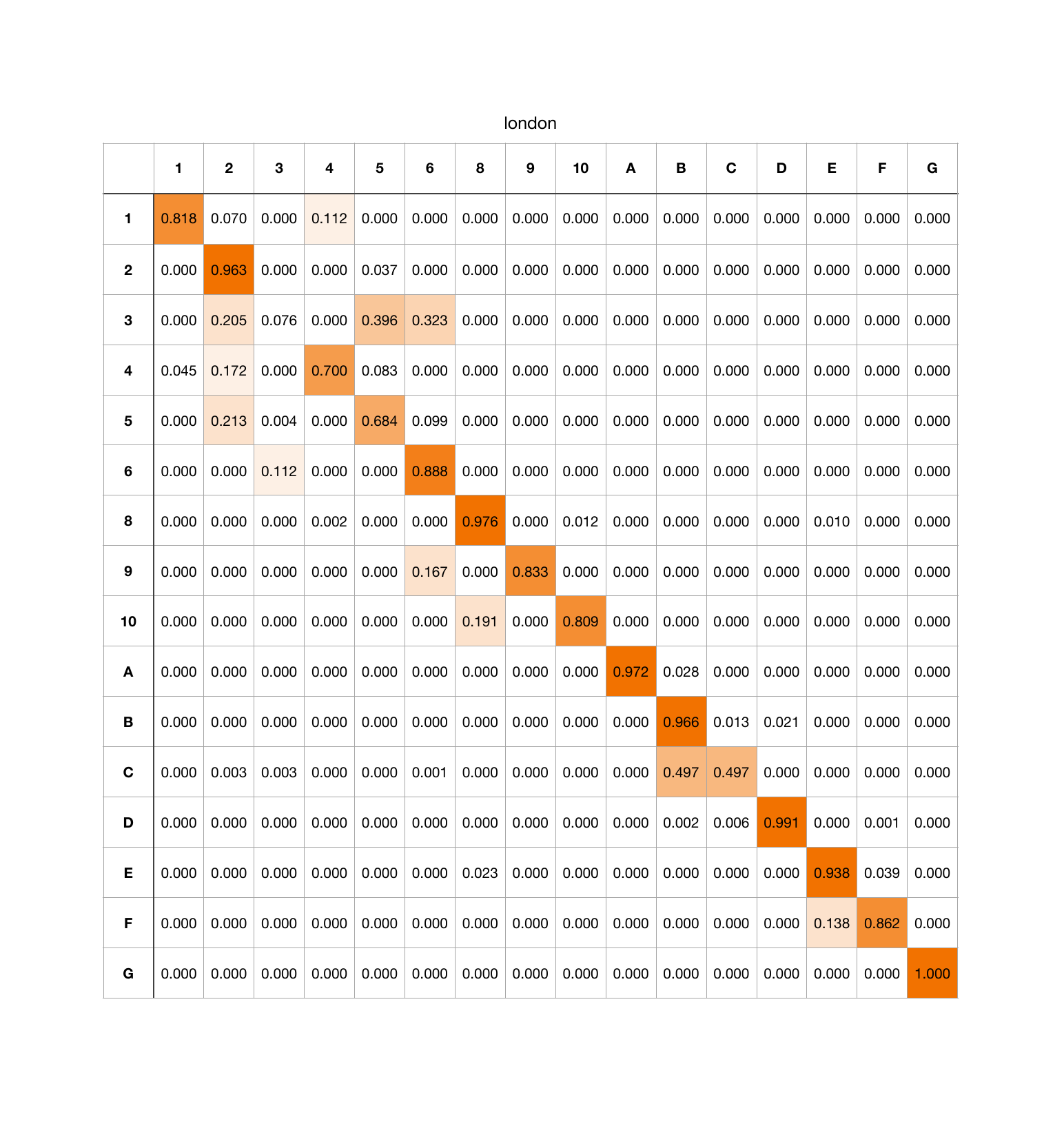}
    \includegraphics[width = 0.494\textwidth]{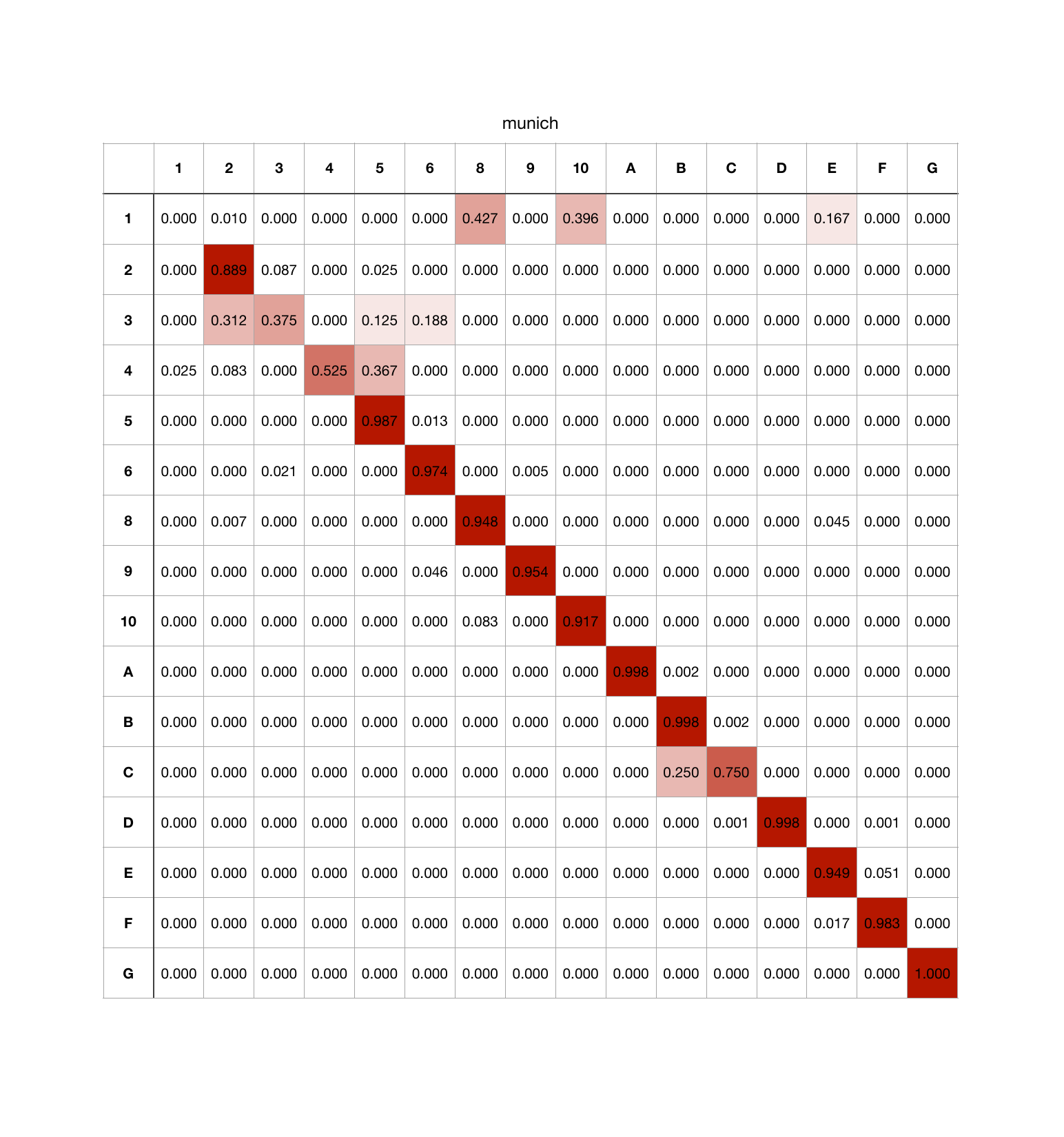}
    \caption[Estimated true confusion matrices for Berlin, London and Munich.]{The matrices shows the true confusion of the voted local climate zones (columns) with the true classes (rows), for Berlin (top left matrix), London (top right matrix) and Munich (bottom matrix).}
   \label{fig:theta_cities}
 \end{figure}

Following this idea, one assumes that the distribution of $Z$ in different cities can differ, i.e.\ the parameters $\pi_1,...,\pi_K$ of the multinomial model are different for different regions. The model should be able to represent those differences in terms of different estimated posterior probabilities. But the crucial aspect here are the misclassification probabilities matrices $\Theta$. Does the voting behaviour of experts look different in different regions? In order to answer this question, we will further investigate the differences in voting probabilities in various locations. 
This can be achieved by calculating the matrix for all regions separately:
$$ \Theta(s) \mbox{ for } s=\{1,...,S\}. $$
To illustrate the problem, we focus on three regions first: Berlin, London and Munich. Figure \ref{fig:theta_cities} shows these regional confusion matrices $\Theta_B, \Theta_L$ and $\Theta_M$ for the example cities. The values on the diagonal refer to the probability of correctly classifying an image to its corresponding LCZ. The three cities were chosen as examples as they differ in their general structure and have a large number of labelled patches. As the plots and the values of the matrices show, the voting probabilities and therefore the probability of correct classification of certain classes are different. This supports the claim that some classes might be more difficult to spot in some cities than in others.

Looking at the general setting and comparing all pairs of cities, we can conduct statistical tests to assess the geographic differences in the voting behaviour of experts. After estimating $\Theta(s)$ and $\Theta(s')$ separately, we want to test the hypothesis $H_0: \Theta(s) = \Theta(s')$. Therefore, we make use of the vectorizations of $\Theta$ and the variance estimates of the parameters, as described in Section 3. The difference between the vectorizations of the confusion matrices is denoted by 
$$ \hat{\boldsymbol{\delta}} = \hat{\boldsymbol{\vartheta}}(s) - \hat{\boldsymbol{\vartheta}}(s'), $$
with $E(\hat{\boldsymbol{\delta}})$ = 0 if $H_0$ holds. Additionally, we know that $Var(\hat{\boldsymbol{\delta}}) = Var(\hat{\boldsymbol{\vartheta}}(s))+ Var(\hat{\boldsymbol{\vartheta}}(s'))$. 
The test statistic is constructed as
$$ T = Var(\hat{\boldsymbol{\delta}})^{-1/2} \hat{\boldsymbol{\delta}}, $$ 
where $Var(\hat{\boldsymbol{\delta}})^{-1/2}$ is calculated by using a singular value decomposition. Under $H_0$ it holds 
$$ T \overset{a}{\sim} N(0, \boldsymbol{I}),$$
which suggests using ordinary one-sample t-tests to test the null hypothesis of equal confusion matrices for cities $s$ and $s'$. 
Repeating this procedure for all pairs of cities leads to the p-values reported in Table \ref{tab:city_pvals}. On a significant level of 0.05, we can assume that the confusion matrices are different for 11 of 34 city pairs.
Returning to the exemplary matrices in Figure \ref{fig:theta_cities}, we conclude that $\Theta_{B}$ is significantly different from $\Theta_{L}$, but not from $\Theta_{M}$. In other words, we can demonstrate that misclassification probabilities differ in the different cities. \\
It should be noted that we omitted estimation variability in the derivation of the $p$ values above. In other words, we considered the city-specific confusion matrices as fixed. This is to some extent plausible since the sample size is rather large and hence the estimation variance of $\Theta$ is small. 
To confirm this we run a few outer bootstrap loops, which are described and reported in the Supplementary Material.

\begin{table}

\caption{\label{tab:city_pvals}P-values of the pairwise tests for differences between cities.}
\flushleft
\footnotesize
\fbox{%
\begin{tabular}{lccccccccc}
	\toprule
	& cologne & london & madrid & milan & munich & paris & rome & zurich \\ 
	\midrule
	berlin &    0.760 &  0.011 (*) &  0.429 &  0.136 &  0.677 &  0.809 &  0.390 &  0.003 (*) \\
	cologne &      &  0.006 (*) &  0.945 &  0.237 &  0.000 (*)&  0.837 &  0.346 &  0.001 (*)\\
	london &      &     &  0.001 (*)&  0.103 &  0.506 &  0.015 (*)&  0.208 &  0.397 \\
	madrid &      &     &     &  0.065 &  0.106 &  0.938 &  0.470 &  0.004 (*) \\
	milan &      &     &     &     &  0.415 &  0.021 (*)&  0.414 &  0.008 (*) \\
	munich &      &     &     &     &     &  0.112 &  0.902 &  0.740 \\
	paris    &      &     &     &     &     &     &  0.213 &  0.001 (*) \\
	rome &      &     &     &     &     &     &     &  0.116\\
	\bottomrule
\end{tabular}}
\end{table}

\section{Discussion}

The paper demonstrates, that the labelling of images is subject to error, misclassification and heterogeneity of labellers. The results are relevant for all machine learning applications where image classification is pursued on multiply labelled data. Ambiguity is inevitable and the current paper aims to quantify this. It is important to note that error and uncertainty in the labelling process might stem from different sources and is multi-dimensional, as we showed in Section 4. In the context of classifying satellite images into climate zones, we were able to detect three primary sources of label uncertainty. These can be analysed based on the assumption that a latent ground truth label exists, on which the labellers condition their assessment. 

First,  the distinguishability of classes is not equal. This aspect is crucial not only for the earth observation domain, but relevant for most applications of image classification, be it medical image or face recognition. On the one hand, our analysis confirms that urban classes are much more challenging to identify than non-urban classes. On the other hand, the identifiability of classes also depends strongly on the database. Therefore, balanced classes are desirable and could improve and stabilise the labelling procedure. 

Second, a fundamental aspect is labellers' heterogeneity and voting behaviour. Here, special training of the labellers was required as the classification of satellite images is a non-trivial task, even for earth observation experts. As the experts received the same training, one would expect homogeneity. However, we demonstrated an approach to assess homogeneity and found differences between the labellers. These can play a huge role, particularly if the panel of human labellers differ between the images, a problem not occurring in our data by design of the labelling process. Generally, labeller heterogeneity should not be neglected in the analysis of uncertainty. 

Third, external properties of the instances to be classified can impact the labelling accuracy and therefore increase label uncertainty. In our case, the origin city for each image impacted the classification probabilities. While we only analysed this aspect by using separately estimated parameters, one could also include the variable in the model and inspect its impact. We have dispensed with this here, as the resulting model would have required estimating a huge number of parameters. Nevertheless, we note that incorporating external knowledge about the images could presumably lead to improved results. As already indicated in Section 4.3, the cause of the observed differences can not only be the voting behaviour of the experts but also arise due to the pursued strategy of selecting images and polygons for labelling. Imbalanced classes might induce a bias and increase the variance in the results which should be taken into account. This is an important aspect and leads to a number of new questions, related to the topics of survey methodology, sampling theory and active learning, see e.g. \citet{Settles:2009} or more recently \citet{Budd:2021}.
 
As a next step, it would be helpful to include the uncertainty in the machine learning process as well. The labelling process only serves as a preprocessing step for the data at hand and produces labelled training data. This data has been used for building elaborate models and networks to classify satellite images into local climate zones automatically. Therefore, if the training data is flawed or suffers from high label uncertainty already, the same holds for the subsequent models. The results obtained by analysing the sources of labelling uncertainty and being able to quantify them could create possibilities to improve and stabilise machine learning processes in terms of overall uncertainty, a topic apparently beyond the scope of this paper.

\appendix
\section{EM Algorithm}

We look at the (artificial) complete likelihood, resulting when $Z^{(i)}$ is known, i.e. the true image class is given. In this case, the complete log-likelihood results by assuming independence among the images and voters as
\begin{align*}
	\log p(Y, Z;\boldsymbol{\pi}, \Theta) & = \sum_{i=1}^n \log p(\boldsymbol{Y}^{(i)}, \boldsymbol{Z}^{(i)}; \boldsymbol{\pi}, \boldsymbol{\theta}) = \sum_{i=1}^n \log p(\boldsymbol{Z}^{(i)},\boldsymbol{\pi}) + \sum_{i=1}^n \log p(\boldsymbol{Y}^{(i)}|\boldsymbol{Z}^{(i)};\boldsymbol{\theta}_{Z^{(i)}}) \\
	&= \sum_{i=1}^n \log \pi_{Z^{(i)}} + \sum_{i=1}^n \sum_{k=1}^K Y_{k}^{(i)} \log \theta_{Z^{(i)}k}  + \mbox{ parameter-free terms} \\
	&= \sum_{i=1}^n \sum_{l=1}^K \mathbb{1}_{\{Z^{(i)}=l\}} \log \pi_l + \sum_{i=1}^n \sum_{k=1}^K \sum_{l=1}^K  \mathbb{1}_{\{Z^{(i)} = l\}} Y_{k}^{(i)} \log \theta_{lk} + \mbox{ parameter-free terms}. 
\end{align*}

This is a fairly simple model and could easily be estimated by Maximum Likelihood leading to the estimates
\begin{align*}
	\hat{\pi}_l = \frac{1}{n} \sum_{i=1}^n \mathbb{1}\{Z^{(i)} = l\}; \quad
	\hat{\theta}_{lk} = 
	\frac{ \displaystyle \sum_{i=1}^n \mathbb{1}\{Z^{(i)} = l\}
		\mathbb{1}\{Y^{(i)} = k\}}
	{     \displaystyle  \sum_{i=1}^n \mathbb{1}\{Z^{(i)} = l\}}
\end{align*}
Apparently, the likelihood above assumes that  the true image class is given in the data. This is not the case, which brings us to the  popular estimation strategy of the EM algorithm, described below.

\section{Label Switching in SEM}
Put into an algorithmic format, we get the following procedure: 
\begin{enumerate}
	\item For $L=\{L_1,...,L_K\}$ and $C=\{C_1,...,C_K\}$:
	\begin{enumerate}
		\item Sort $L$ in descending order according to the relative frequency of the labellers' votes
		\item Apply the permutation to all $l=1,...,K$, s.t.: $\sigma^{-1}(k) = \mbox{arg max}_{l}\big(P( Z^{(i)} = l | V^{(i)} = k )\big) $
	\end{enumerate}
	\item If a relation $k \rightarrow l$ found in step (b) is unique, delete $C_k$ and $L_l$ from $C$ and $L$.
	\item Repeat until a unique allocation $C \rightarrow L$ is found.
\end{enumerate}

\section{Location of images and polygons}

\begin{figure}[h]
	\centering 
	Berlin \hspace{2.5cm} Cologne \hspace{2.5cm} London\\
	\includegraphics[width=0.25\textwidth]{Plots/Maps/berlin_all.png}
	\includegraphics[width=0.25\textwidth]{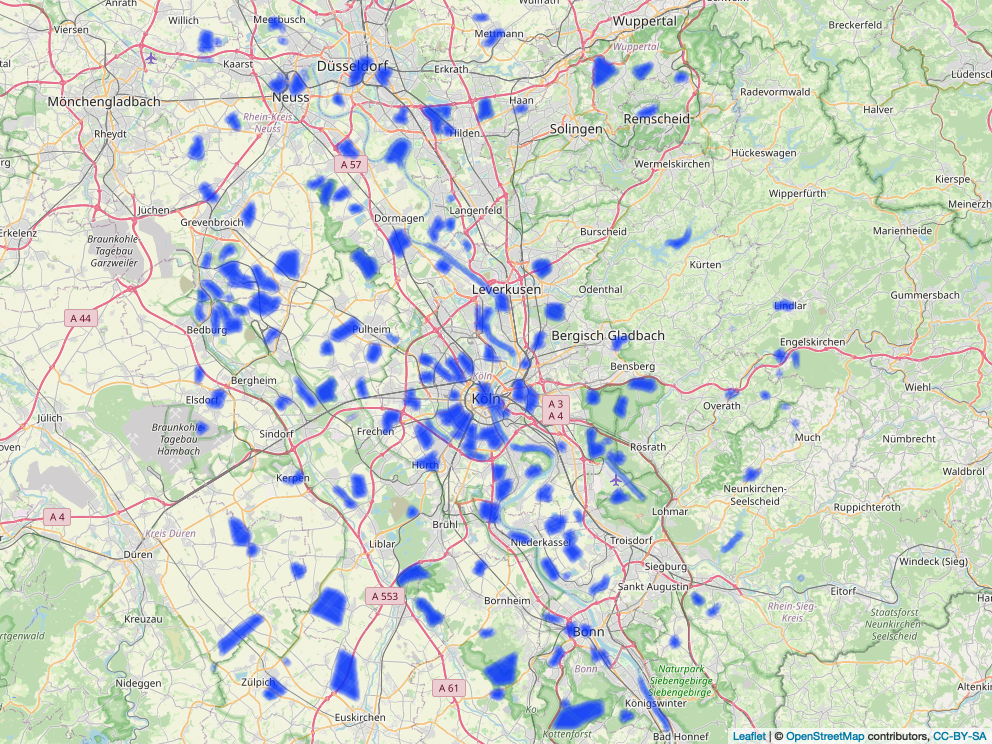}
	\includegraphics[width=0.25\textwidth]{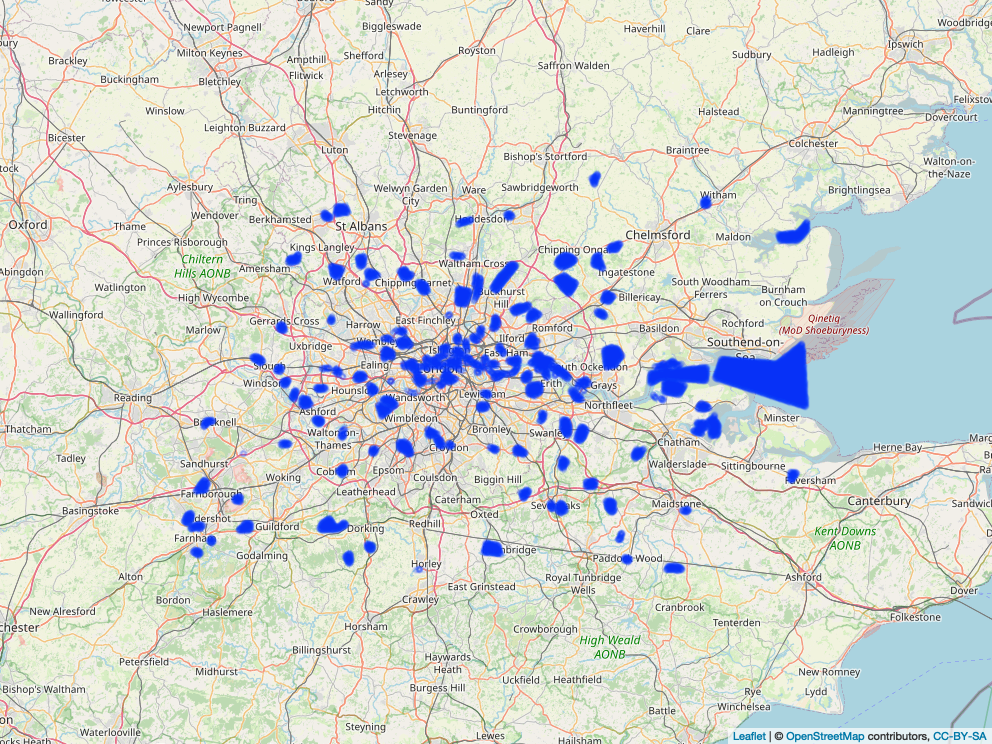}
	\\
	Madrid \hspace{2.5cm} Milan \hspace{2.5cm} Munich \\
	\includegraphics[width=0.25\textwidth]{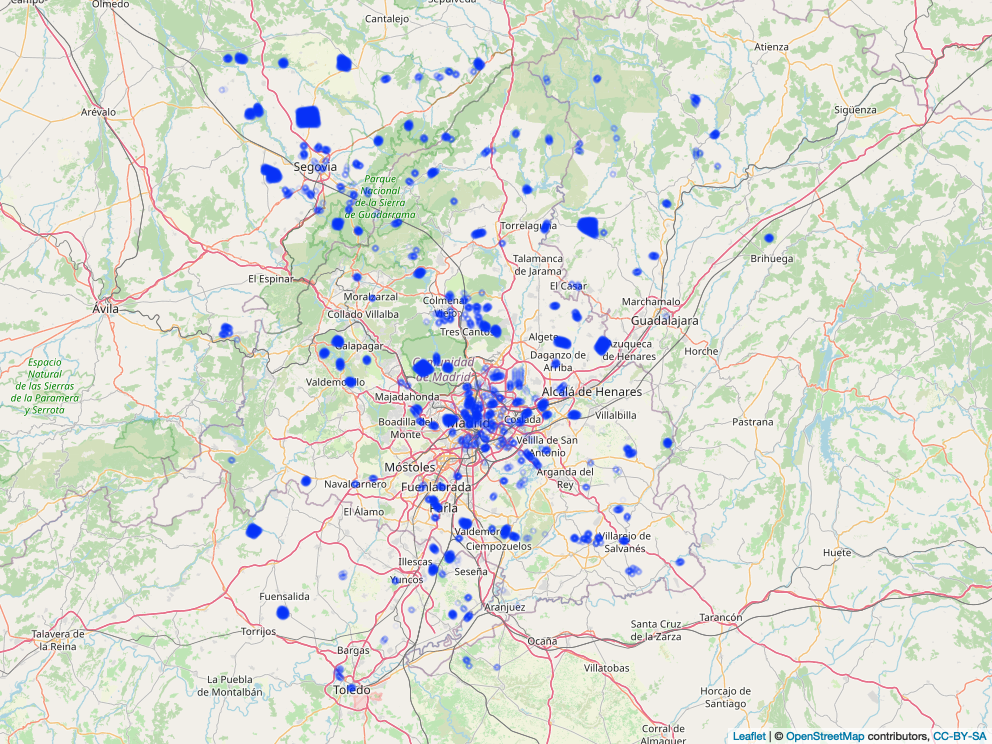}
	\includegraphics[width=0.25\textwidth]{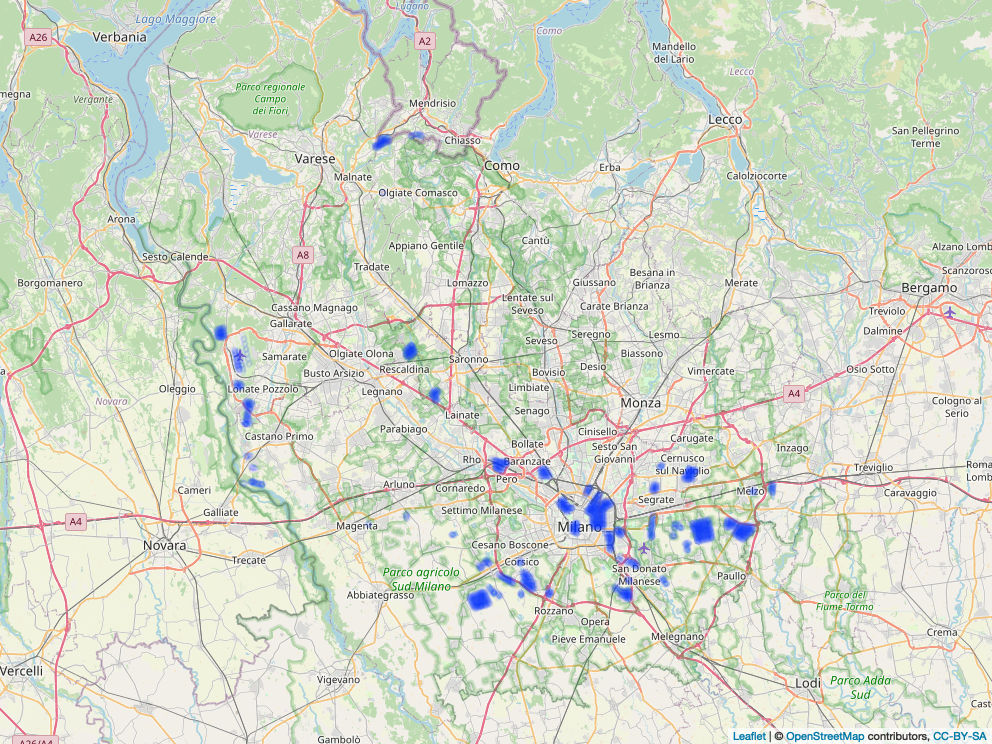}
	\includegraphics[width=0.25\textwidth]{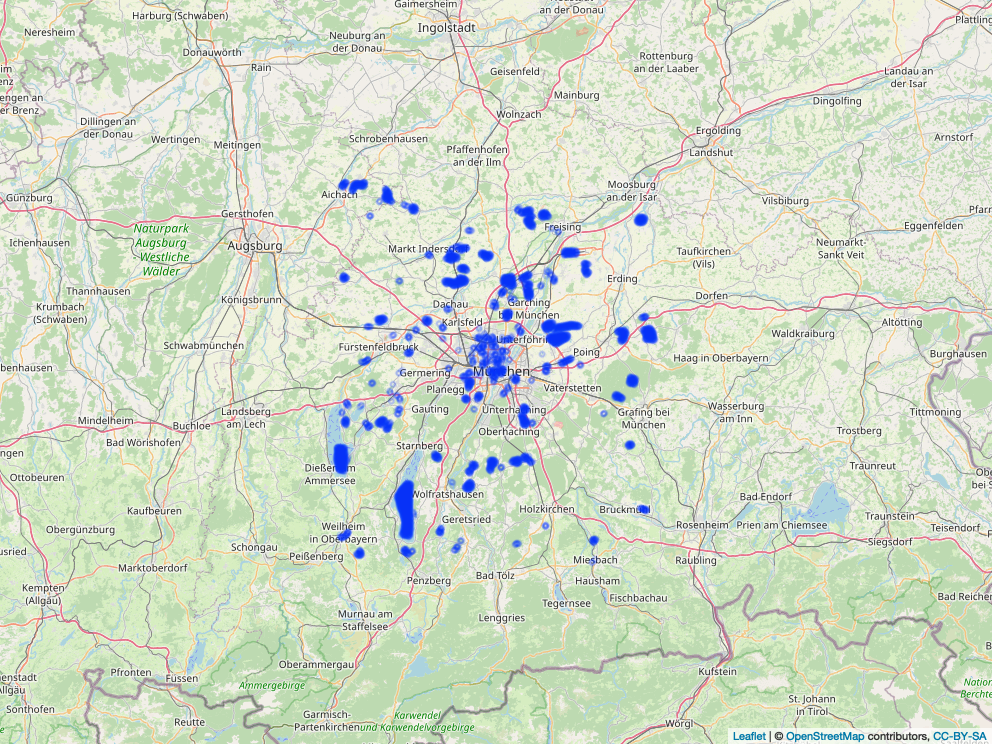} 
	\\
	Paris \hspace{2.5cm} Rome \hspace{2.5cm} Zurich \\
	\includegraphics[width=0.25\textwidth]{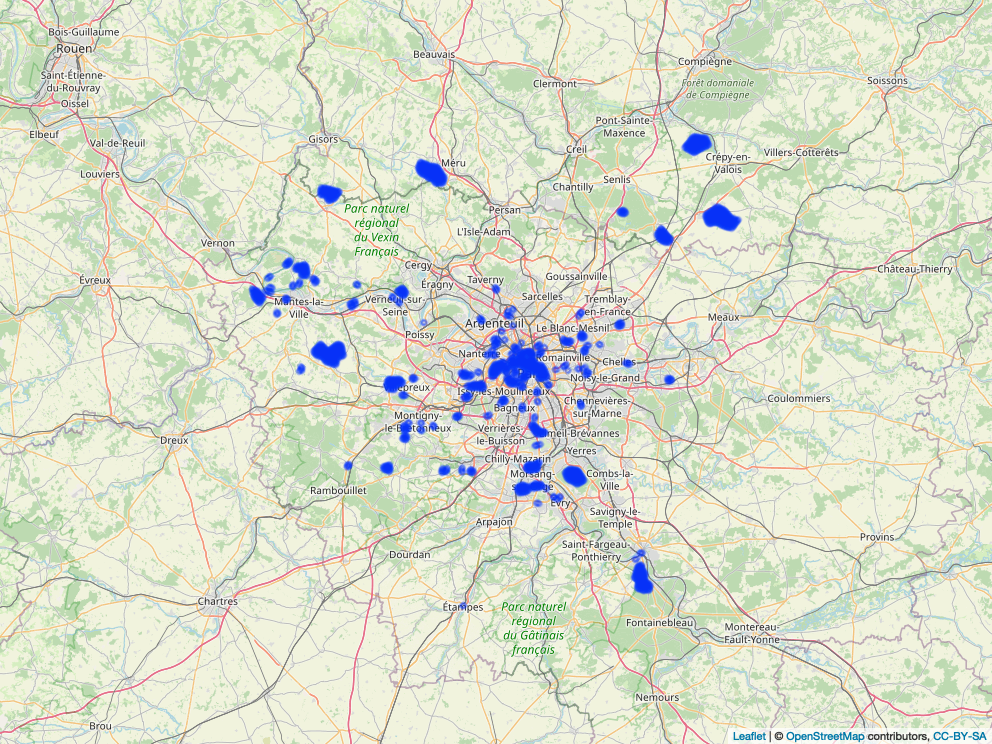}
	\includegraphics[width=0.25\textwidth]{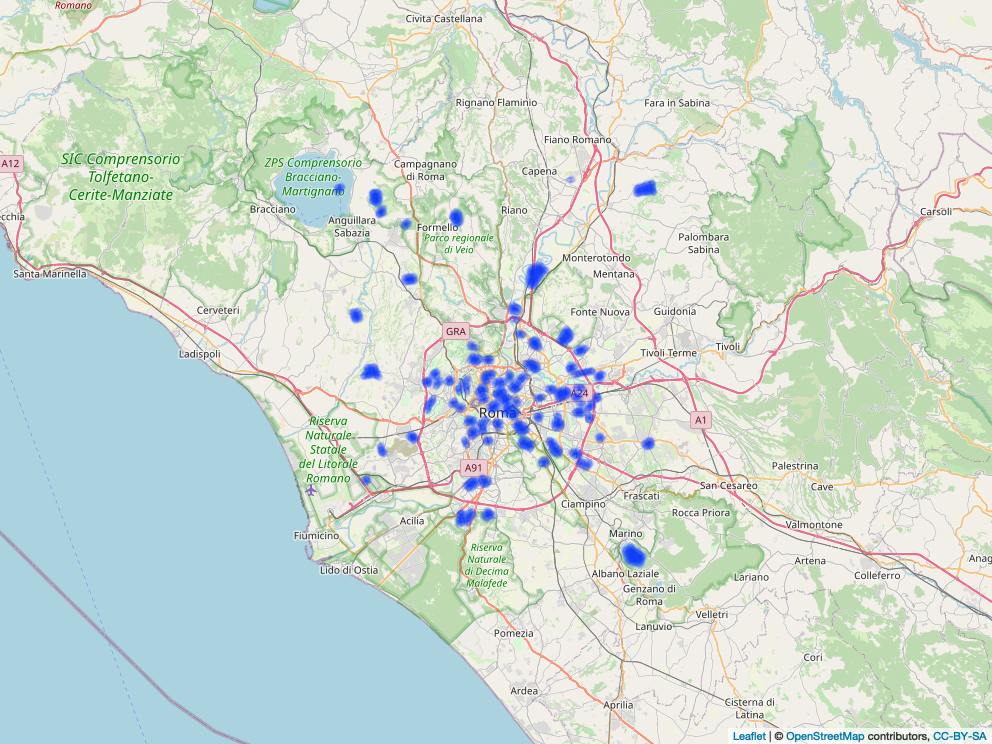}
	\includegraphics[width=0.25\textwidth]{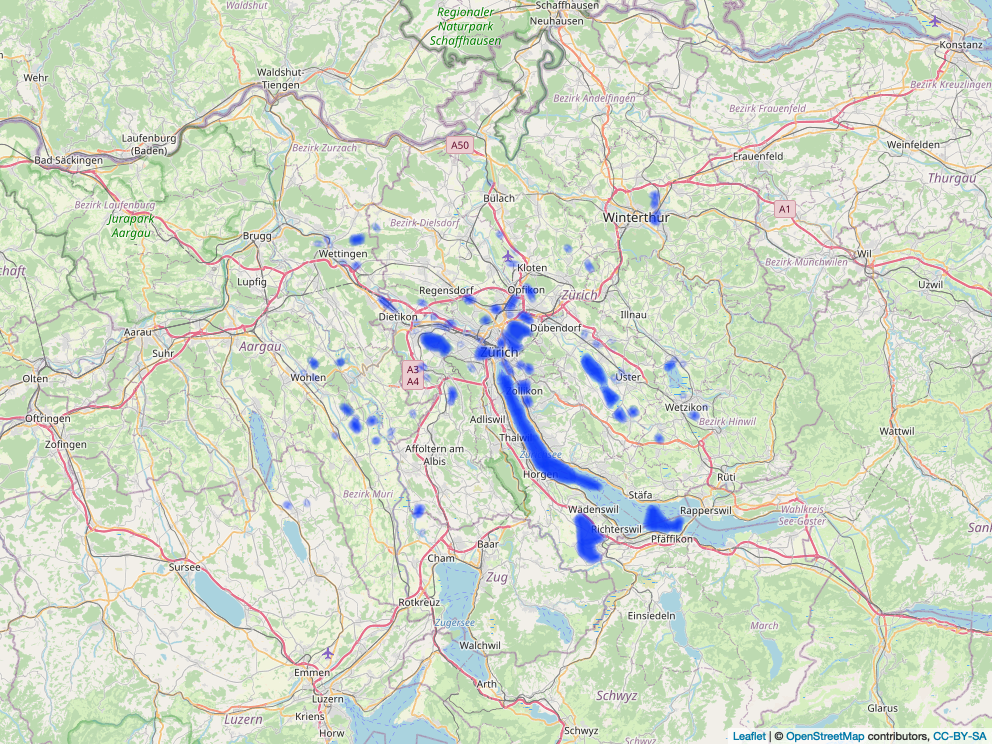}
	
	\caption{The figure shows the maps of all cities with the coordinates of the images indicated as blue spots. }
	\label{fig:Maps_Locations}
\end{figure}

As mentioned in Section 4.3, the observed differences between the cities might not be caused solely by different voting behaviors. The pursued sampling strategy and selection of images and polygons possibly lead to imbalanced labels and therefore bias in the observed results. Looking at the coordinates of the images and their distribution across the cities, as shown in Figure \ref{fig:Maps_Locations} supports this hypothesis.

\section*{Acknowledgements}
The present contribution is supported by the Helmholtz Association under the joint research school “HIDSS-006 - Munich School for Data Science@Helmholtz, TUM\&LMU.

\bibliography{literature}

\end{document}


\maketitle

\section{Estimation Uncertainty via Bootstrap}

The analyses and especially $p$ values derived in the paper rely on the estimates and estimation variability is not taken into account. To do so we run a few outer loop bootstrap simulations and re-estimate the parameters to conduct the analyses and tests multiple times. This section explains the details and provides the results of the bootstrapped  testing procedures. 

\subsection{Expert Heterogeneity}
We bootstrap the dataset $Y^{(i),(*b)}$ for $b=1, \ldots , B$ by randomly drawing with replacement (!) from the set of all observation $Y^{(i)}$ for $i \in {\cal I}$. We then re-run the fitting procedure on the bootstrapped datasets. This results in bootstrapped estimates $\hat{\pi}_k^{(*b)}$, $\hat{\Theta}^{(*b)}$ and in particular $\hat{\tau}^{(*b)}$, as well as $\hat{\tau}_{(-j)}^{(*b)}$ (posterior probabilities for each expert excluded) from which we can simulate the p-value as described above. 
While we experience stable estimation due to the large  number of available images, the bootstrapped testing results do vary for some singular bootstrap iterations. This occurred due to problems with our label switching strategy, as the labels for non-frequently voted classes are sometimes not consistent with the estimated data and hence lead to varying results. We, therefore, need to consider label switching in the bootstrap runs as well and hence introduce a second relabelling  to match the classes of the bootstrapped estimates to the original estimates. Similar to the first strategy described in Section 3 and Appendix B, we calculated the posterior probabilities of the latent classes 
$$P = diag(\pi) \Theta^T$$
for both the original estimates and its bootstrapped versions. The rows of the resulting matrices refer to the estimated true classes and are therefore subject to label switching. In order to get a proper matching, we matched each row of the bootstrapped estimate to the most similar row of the original estimate and shuffled the rows accordingly. Denoting the original class as $k$ and the bootstrapped class as $l$, we can formulate the permutation rule
$$ \sigma^{-1}(k) = arg \min_l (P_k - P^{(*b)}_l)^2. $$
This permutation is applied to the classes in the descending order of the size of the original estimates $\hat{\pi}_k$ and if no unique matching is found in the first run, the procedure is repeated for all classes that are not allocated yet. 
This procedure leads to reasonable consistent estimates and is necessary to compare the results of the different bootstrap iterations. 

\begin{figure}
    \centering
    \includegraphics[width=0.6\textwidth]{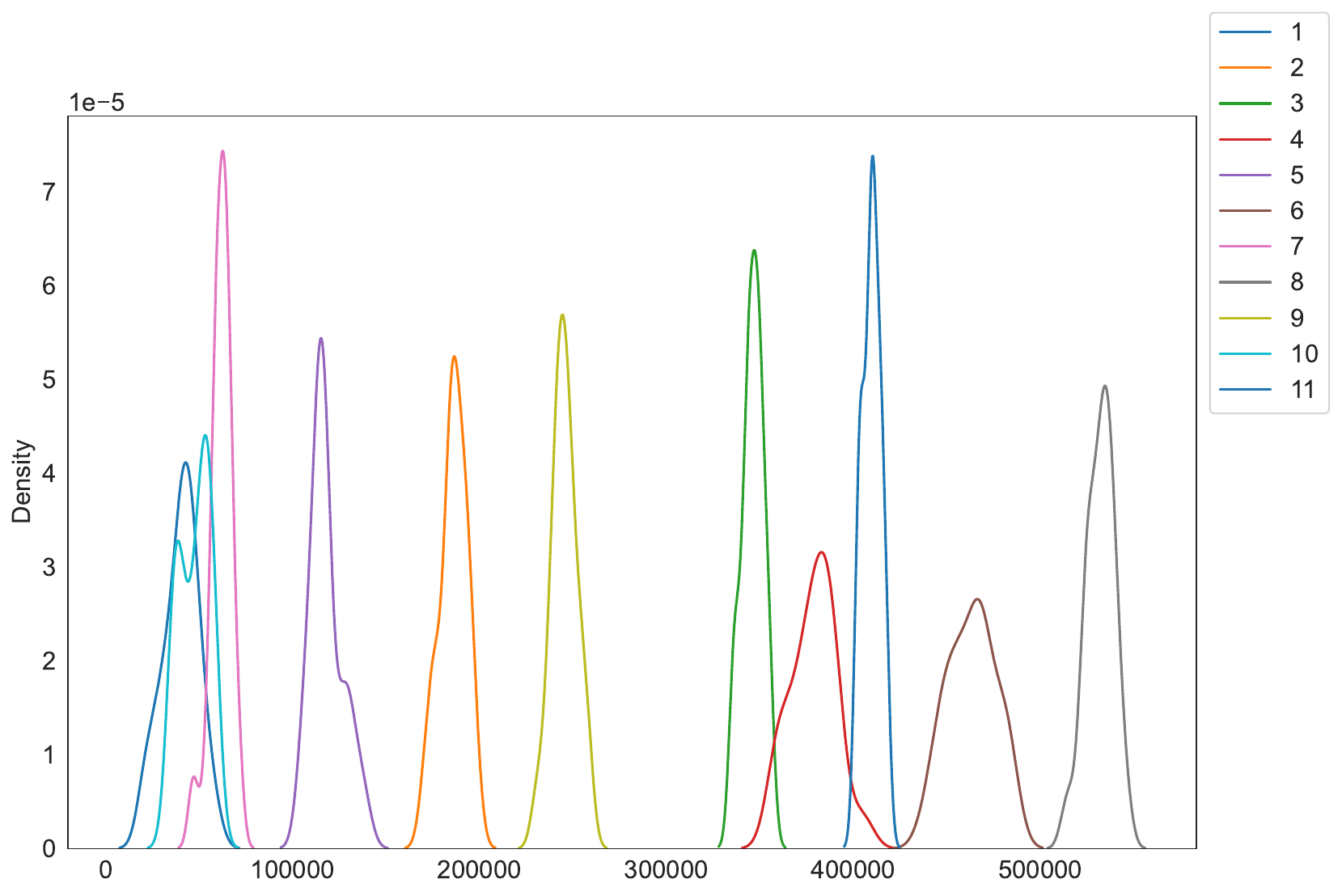} \\
    \includegraphics[width=0.6\textwidth]{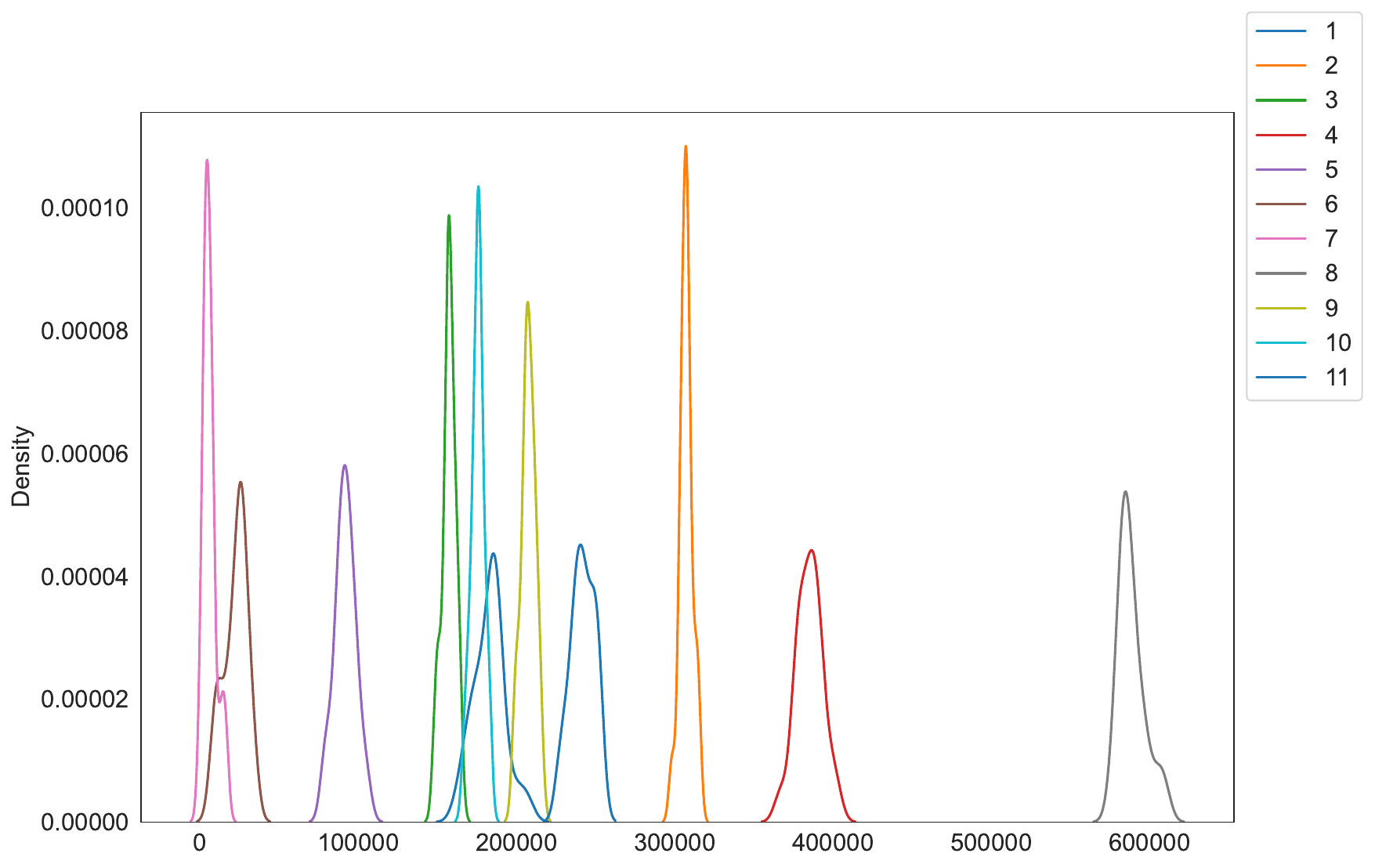} \\
    \includegraphics[width=0.6\textwidth]{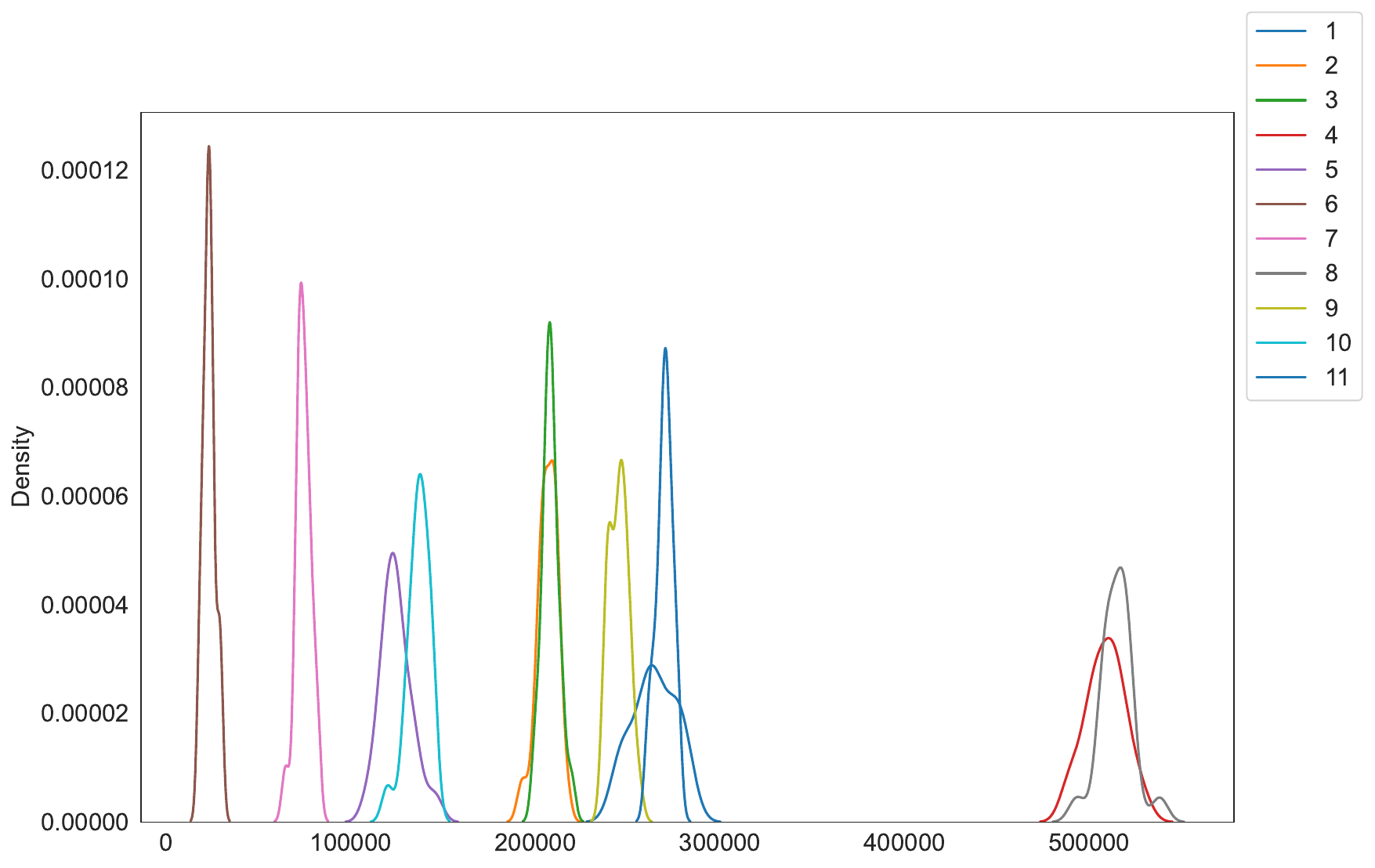}
    \caption{Densities of differences of goodness-of-fit statistics per expert, for three bootstrap iterations.}
    \label{fig:bootstrap_expert_density}
\end{figure}

\begin{figure}
    \centering
    \includegraphics[width=0.6\textwidth]{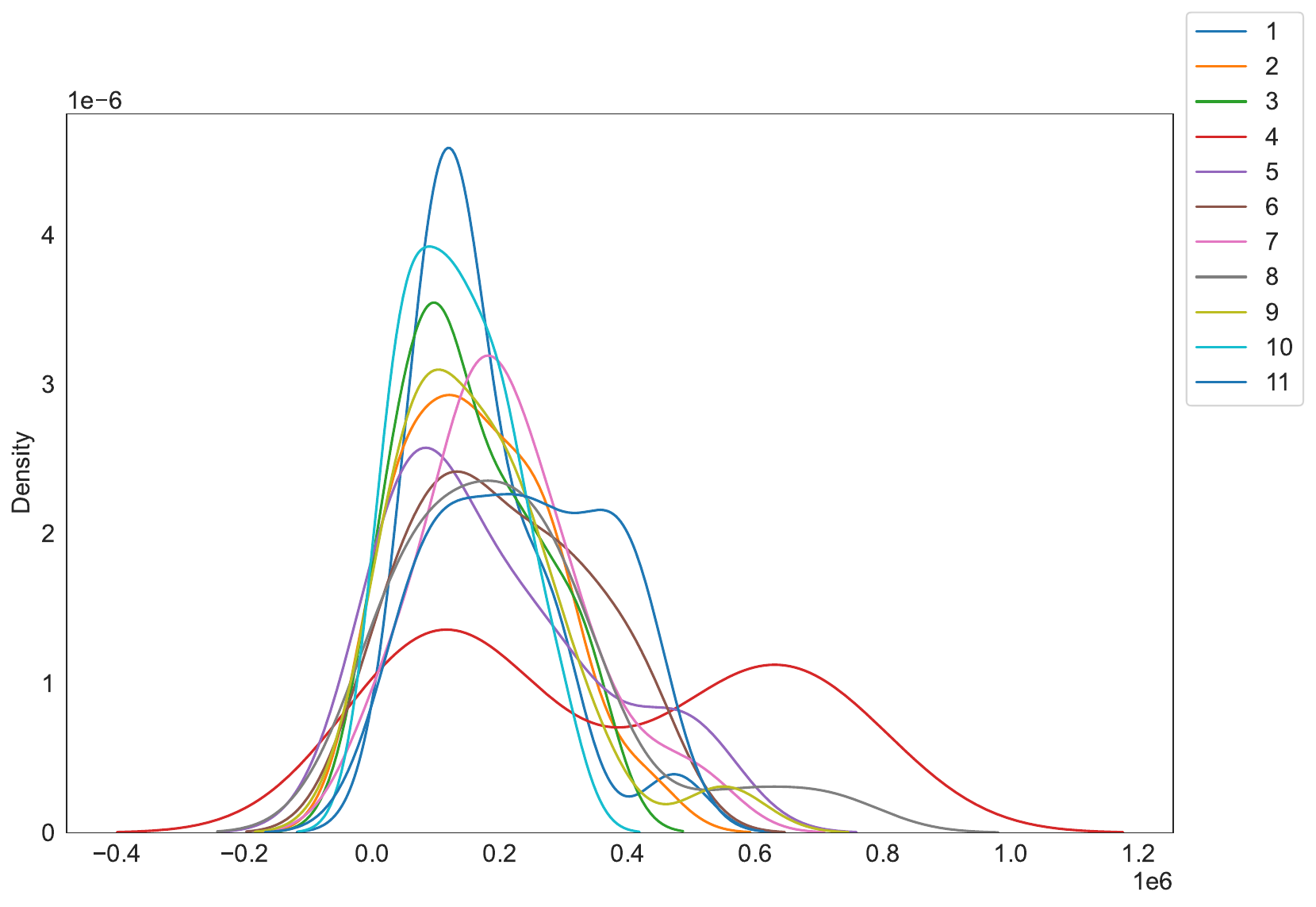} \\
    \includegraphics[width=0.6\textwidth]{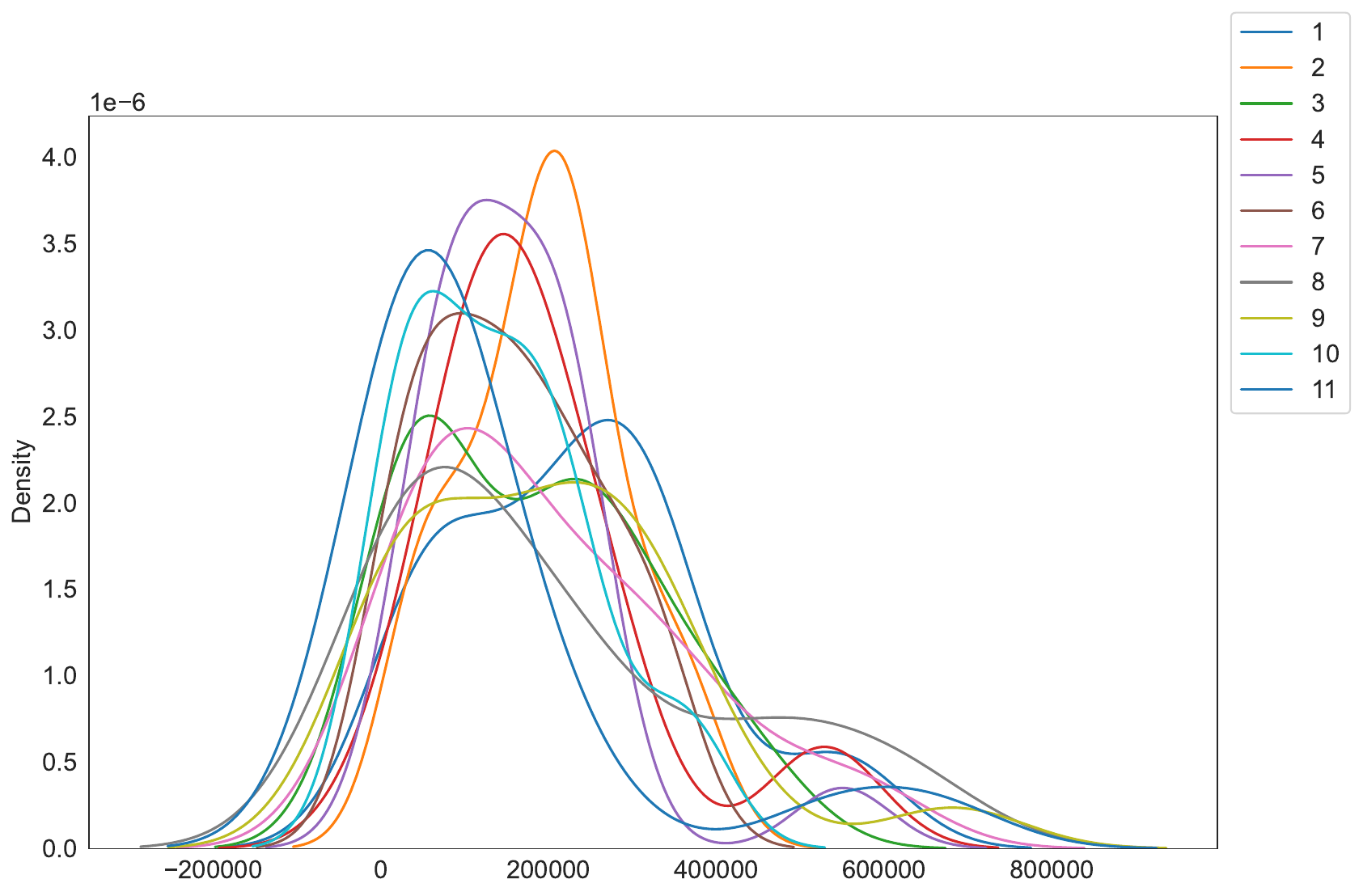} \\
    \includegraphics[width=0.6\textwidth]{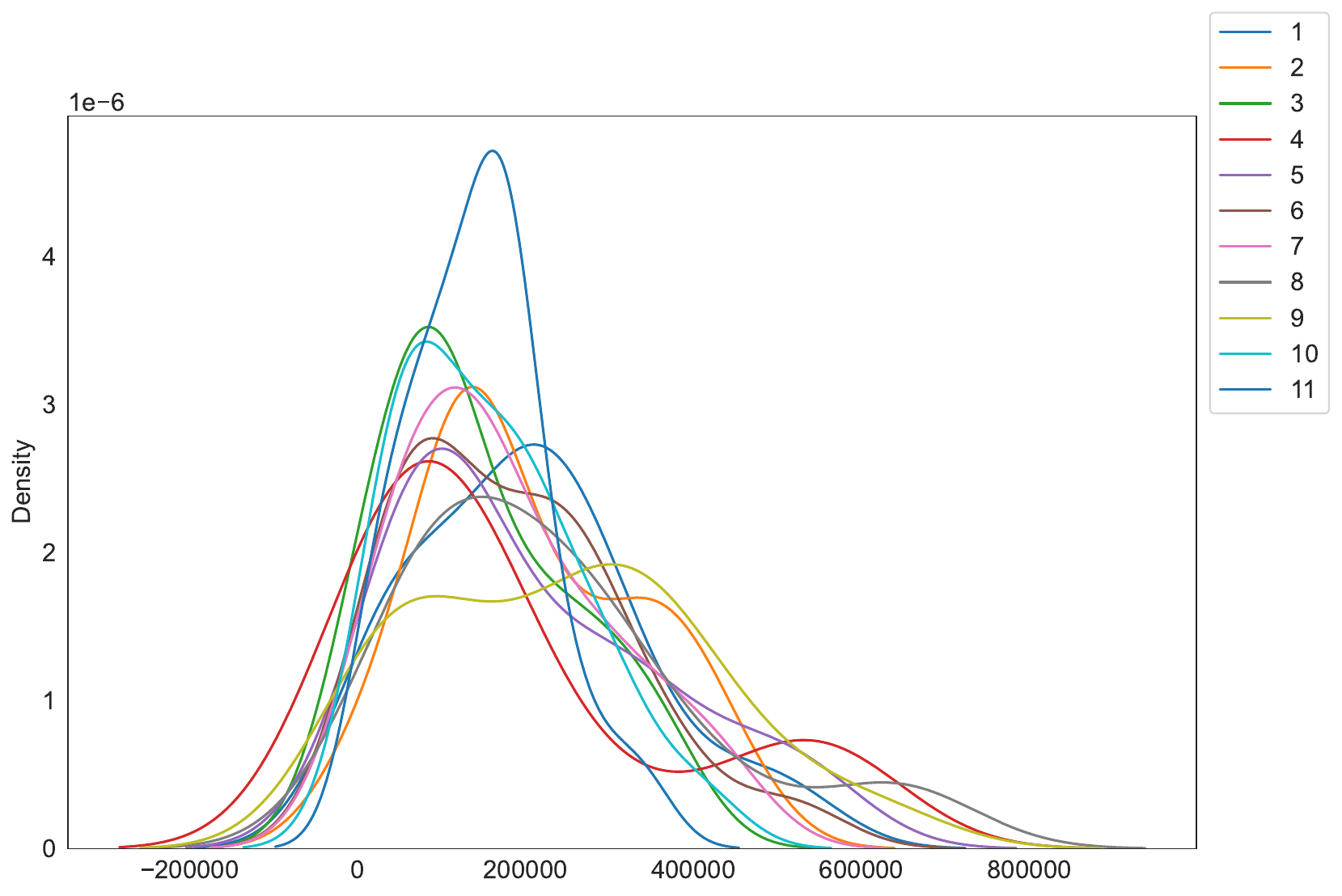}
    \caption{Densities of differences of the randomized goodness-of-fit statistics per expert, for three bootstrap iterations.}
    \label{fig:bootstrap_expert_density_random}
\end{figure}

\begin{figure}
    \centering
    \includegraphics[width=0.45\textwidth]{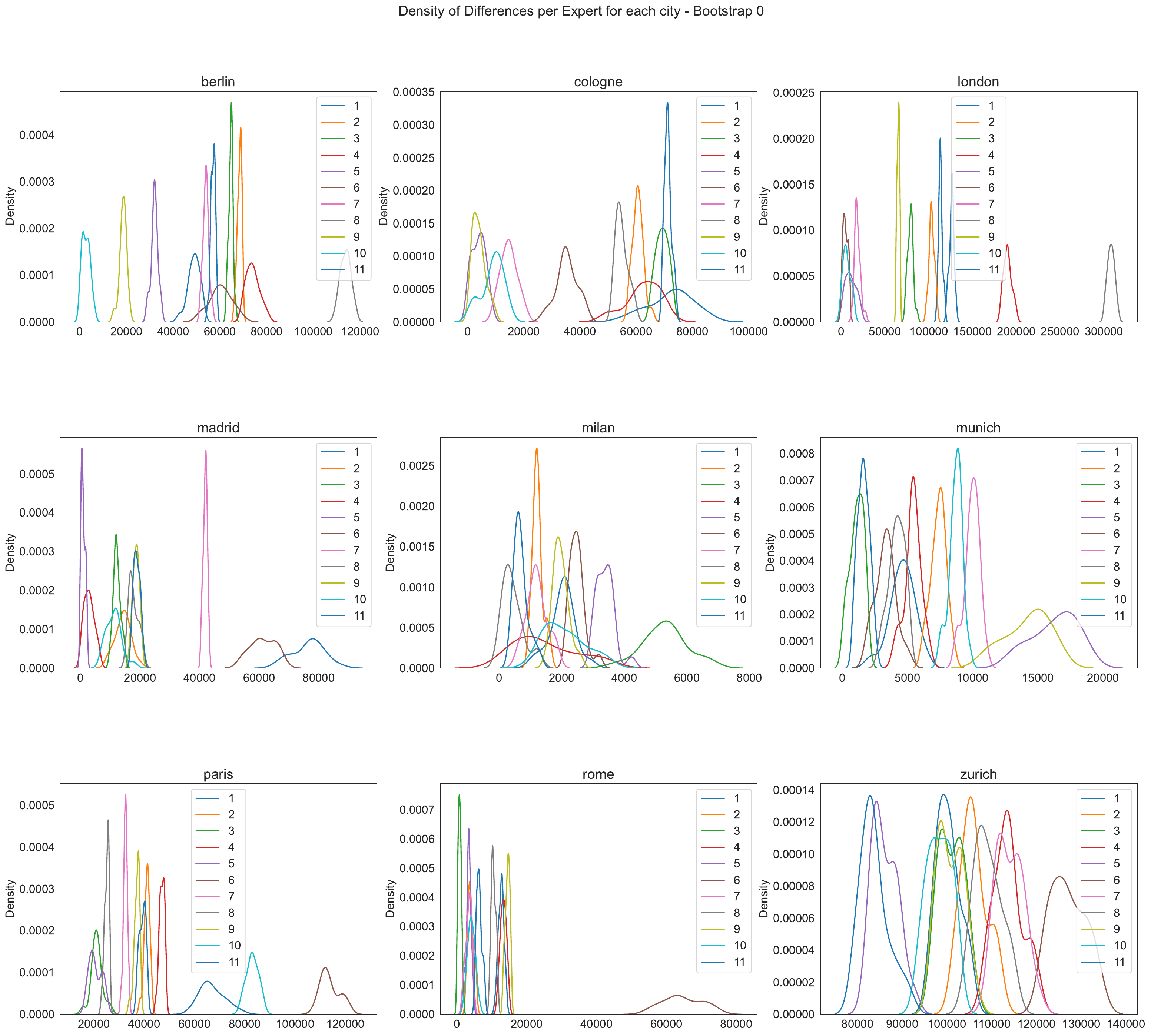} 
    \includegraphics[width=0.45\textwidth]{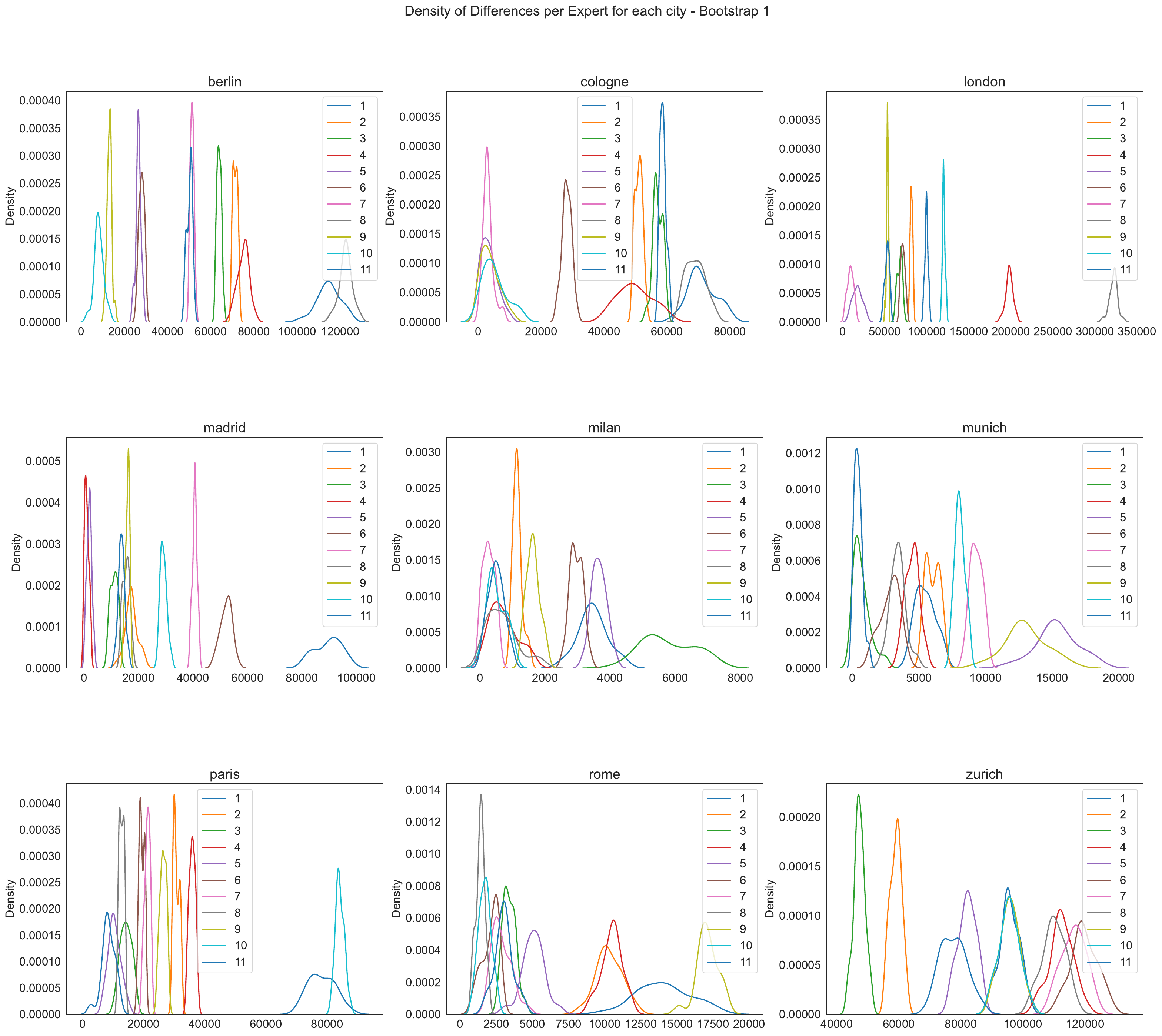} \\
    \includegraphics[width=0.45\textwidth]{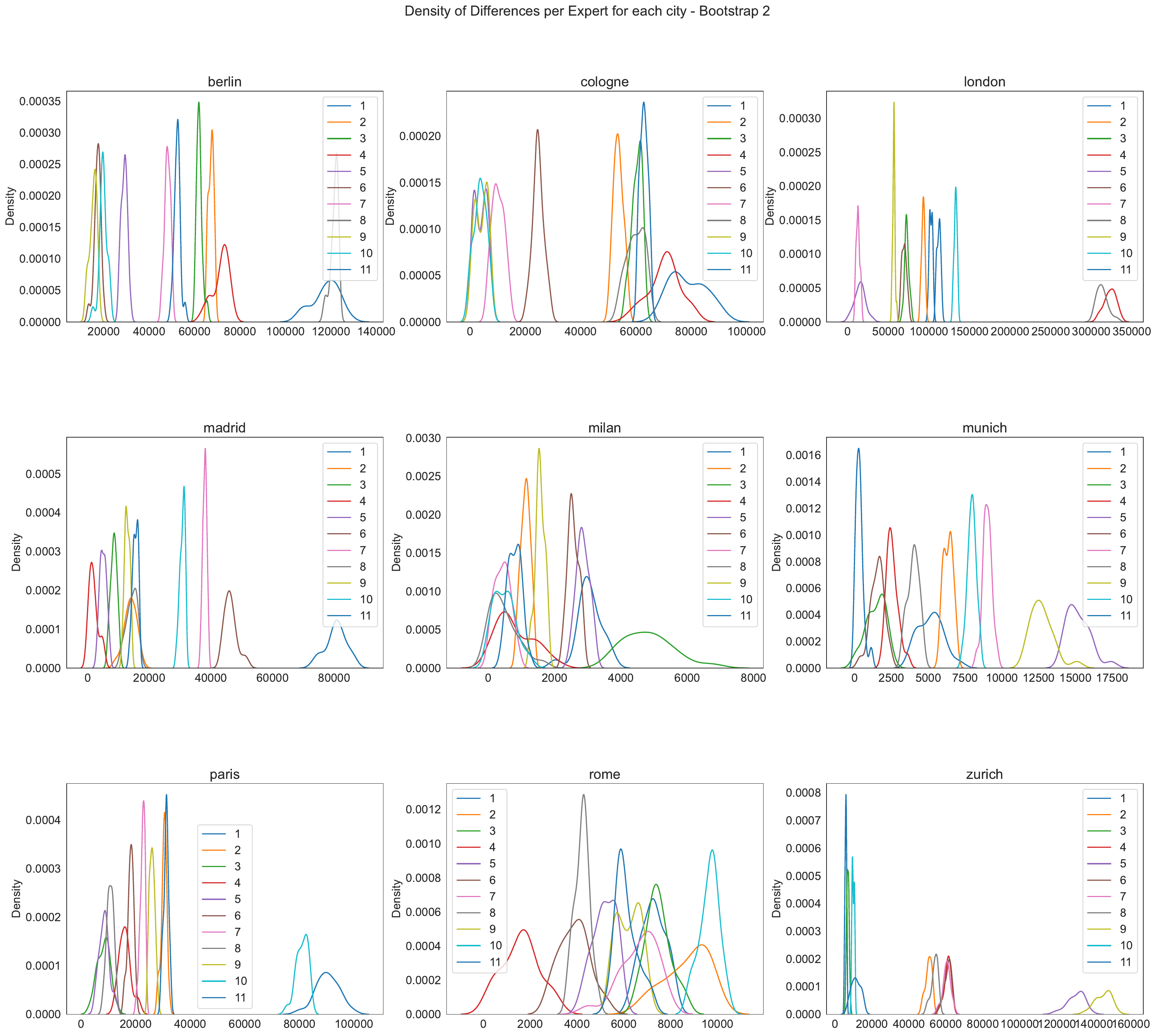} 
    \label{fig:bootstrap_expert_density_cities}
    \caption{Densities of the goodness-of-fit statistics per expert, per city, for the three bootstrap iteraions.}
\end{figure}

For analysing the expert heterogeneity, we did not conduct an explicit significance test. Nevertheless, it is worth taking a look at the results of a few bootstrap iterations. Following the described strategy, we draw a bootstrap sample and obtain bootstrapped estimates of the expert specific posterior probabilities, which can then be used to conduct the analyses described in Section 4.2. Figure \ref{fig:bootstrap_expert_density} shows the densities of the differences of the goodness-of-fit statistics, calculated for three (outer) bootstrap iterations. These plots show that there might be some variation in the estimated parameters but the tendencies of some experts towards a heterogeneous voting behaviour remain nevertheless. \\
Again, we randomized the procedure and calculated the same quantities with random votes, leading to the plots shown in Figure \ref{fig:bootstrap_expert_density_random}. \\
The plots in Figure \ref{fig:bootstrap_expert_density_cities} show results for the split by the different cities. We see that the voting behaviour remains roughly stable for cities with a larger sample size (e.g. Berlin, London or Cologne) but varies quite a lot for cities with fewer images (e.g. Zurich, Rome or Milan). This is reasonable and again shows that stable estimation is always subject to a sufficient data basis.

\newgeometry{left=5mm, right=5mm}     
\restoregeometry

\subsection{City Differences} 

For the second testing scenario, let $Y^{(i),(*b)}$ for $b=1, \ldots , B$ be randomly drawn with replacement (!) from the set of observation $Y^{(i)}$ for $i \in {\cal I}$ with ${\cal I}$ as subset by taking images from a particular city labelled as $s \in \{1, \ldots ,S\}$, only. 
We re-run the fitting procedure for each city $s$ separately, based on the bootstrap samples and produce $\hat{\Theta}^{(*b)} (s), s=1,...,S$ as the estimates of interest. We also apply the permutation suggested in the previous section. In this case we are faced with an additional problem. Since only images from the respective city are considered for the bootstrap sample, this leads to rather small sample sizes for some of the cities and classes. 
Nevertheless, we run $B=50$ outer bootstrap loops and test pairwise differences in each of them. Table \ref{tab:city_pvals_bootstrap} shows the p-values for the original estimates, as well as the minimum and median of the bootstrap runs. Originally, we would reject the hypothesis of equal confusion matrices for 11 pairs. For all these pairs, the median p-value is also less than 0.05 and in most cases, even smaller. Hence, we see even when considering the variability of the estimation, that the results above remain valid in that cities are not labelled homogeneously. We also see that the distribution is quite skewed as the median is often quite very close to the minimum value.

\begin{table}
\footnotesize
\centering
\begin{tabular}{llllrr}
\hline
{} &       City 1 &       City 2 &    original p &  minimal p &   median p \\
\hline
0  &   berlin &  cologne &  0.760 &    0.0 &  0.5860 \\
1  &   berlin &   london &  0.011 (*)&    0.0 &  0.0000 \\
2  &   berlin &   madrid &  0.429 &    0.0 &  0.4580 \\
3  &   berlin &    milan &  0.136 &    0.0 &  0.1680 \\
4  &   berlin &   munich &  0.677 &    0.0 &  0.0305 \\
5  &   berlin &    paris &  0.809 &    0.0 &  0.3530 \\
6  &   berlin &     rome &  0.390 &    0.0 &  0.2065 \\
7  &   berlin &   zurich &  0.003 (*)&    0.0 &  0.0000 \\
8  &  cologne &   london &  0.006 (*)&    0.0 &  0.0000 \\
9  &  cologne &   madrid &  0.945 &    0.0 &  0.6680 \\
10 &  cologne &    milan &  0.237 &    0.0 &  0.1280 \\
11 &  cologne &   munich &  0.000 (*)&    0.0 &  0.0145 \\
12 &  cologne &    paris &  0.837 &    0.0 &  0.3975 \\
13 &  cologne &     rome &  0.346 &    0.0 &  0.1945 \\
14 &  cologne &   zurich &  0.001 (*)&    0.0 &  0.0000 \\
15 &   london &   madrid &  0.001 (*)&    0.0 &  0.0000 \\
16 &   london &    milan &  0.103 &    0.0 &  0.0390 \\
17 &   london &   munich &  0.506 &    0.0 &  0.1820 \\
18 &   london &    paris &  0.015 (*)&    0.0 &  0.0000 \\
19 &   london &     rome &  0.208 &    0.0 &  0.0080 \\
20 &   london &   zurich &  0.397 &    0.0 &  0.0755 \\
21 &   madrid &    milan &  0.065 &    0.0 &  0.1335 \\
22 &   madrid &   munich &  0.106 &    0.0 &  0.0120 \\
23 &   madrid &    paris &  0.938 &    0.0 &  0.3430 \\
24 &   madrid &     rome &  0.470 &    0.0 &  0.0590 \\
25 &   madrid &   zurich &  0.004 (*)&    0.0 &  0.0000 \\
26 &    milan &   munich &  0.415 &    0.0 &  0.1315 \\
27 &    milan &    paris &  0.021 (*)&    0.0 &  0.0450 \\
28 &    milan &     rome &  0.414 &    0.0 &  0.3385 \\
29 &    milan &   zurich &  0.008 (*)&    0.0 &  0.0010 \\
30 &   munich &    paris &  0.112 &    0.0 &  0.0030 \\
31 &   munich &     rome &  0.902 &    0.0 &  0.1415 \\
32 &   munich &   zurich &  0.740 &    0.0 &  0.0245 \\
33 &    paris &     rome &  0.213 &    0.0 &  0.0585 \\
34 &    paris &   zurich &  0.001 (*)&    0.0 &  0.0000 \\
35 &     rome &   zurich &  0.116 &    0.0 &  0.0000 \\
\hline
\end{tabular}

\caption{Bootstrapped p-values of the tests for differences between cities (B=50). }
\label{tab:city_pvals_bootstrap}
\end{table}

\section{Label Uncertainty in different Classification settings}

In the paper, we state that uncertainty and confusion do not occur equally for all classes. While some LCZs are clearly defined, others cannot be strictly distinguished. 
As discussed in Section 2 of the paper, urban classes are much harder to distinguish than non-urban classes. While the human experts almost never confuse urban and non-urban LCZs, the algorithm tends to classify uncertain images into the more certain non-urban classes. This problem arises because of the lack of votes for some classes and the uncertainty in classification stemming from this issue. In the paper we already state that LCZ 7 was excluded due to the lack of votings and the resulting estimation instability. 
In order to examine this aspect further, we repeat the analyses for different sets of $k=1,...,K$. 

\subsection{Binary: "urban" vs. "nonurban"}
In order to account for the rarity of confusion between classes 1,...,10 and A,...,G in reality, we can additionally inspect a binary classification problem here. Based on the majority voting, we assign the label ’urban’ to images with majority vote {1, ..., 10} and the label ’non-urban’ to those with a majority of experts voting for classes {A, ..., G }. Running the same algorithm to find two clusters leads to the binary true confusion matrix with almost perfect classification. The majority of images stem from the non-urban classes, which is also reflected in the prior probabilities. Images stemming from the true class "urban" were correctly classified in 99.7\% of all cases, while nonurban images could be correctly classified by the experts 99.6\% of all cases 
It is obvious that urban and non-urban images are far easier to distinguish not only for the human experts but also for the algorithm.


Concerning the uncertainty due to different experts, we see a very homogeneous voting behaviour for the binary setting as the distinction between urban and nonurban images is fairly easy. Therefore, the homogeneity analysis does not show any differences in terms of voting for any expert. \\
To inspect the differences between the cities, we again conducted the testing procedure described in the paper. The resulting p-values are displayed in table \ref{tab:binary_p}. For 26 city pairs, we can reject the hypothesis of equal confusion matrices. 

\begin{table}
\centering
\small
\begin{tabular}{lll}
\hline
     City 1 &      City 2 &     p \\
\hline
 berlin & cologne & 0.216 \\
 berlin &  london & 0.000 (*) \\
 berlin &  madrid & 0.304 \\
 berlin &   milan & 0.024 (*)\\
 berlin &  munich & 0.013 (*)\\
 berlin &   paris & 0.161 \\
 berlin &    rome & 0.264 \\
 berlin &  zurich & 0.075 \\
cologne &  london & 0.218 \\
cologne &  madrid & 0.623 \\
cologne &   milan & 0.024 (*)\\
cologne &  munich & 0.061 \\
cologne &   paris & 0.028 (*)\\
cologne &    rome & 0.256 \\
cologne &  zurich & 0.117 \\
 london &  madrid & 0.000 (*)\\
 london &   milan & 0.005 (*)\\
 london &  munich & 0.016 (*)\\
 london &   paris & 0.043 (*)\\
 london &    rome & 0.256 \\
 london &  zurich & 0.000 (*)\\
 madrid &   milan & 0.192 \\
 madrid &  munich & 0.075 \\
 madrid &   paris & 0.522 \\
 madrid &    rome & 0.276 \\
 madrid &  zurich & 0.293 \\
  milan &  munich & 0.033 (*)\\
  milan &   paris & 0.819 \\
  milan &    rome & 0.000 (*)\\
  milan &  zurich & 0.000 (*)\\
 munich &   paris & 0.000 (*)\\
 munich &    rome & 0.230 \\
 munich &  zurich & 0.000 (*)\\
  paris &    rome & 0.704 \\
  paris &  zurich & 0.147 \\
   rome &  zurich & 0.127 \\
\hline
\end{tabular}
\caption{P-values from the testing procedure for equal binary confusion matrices for different cities.}
\label{tab:binary_p}
\end{table}

\subsection{Multiclass: urban classes only}
\begin{figure}
    \centering
    \includegraphics[width=0.5\textwidth]{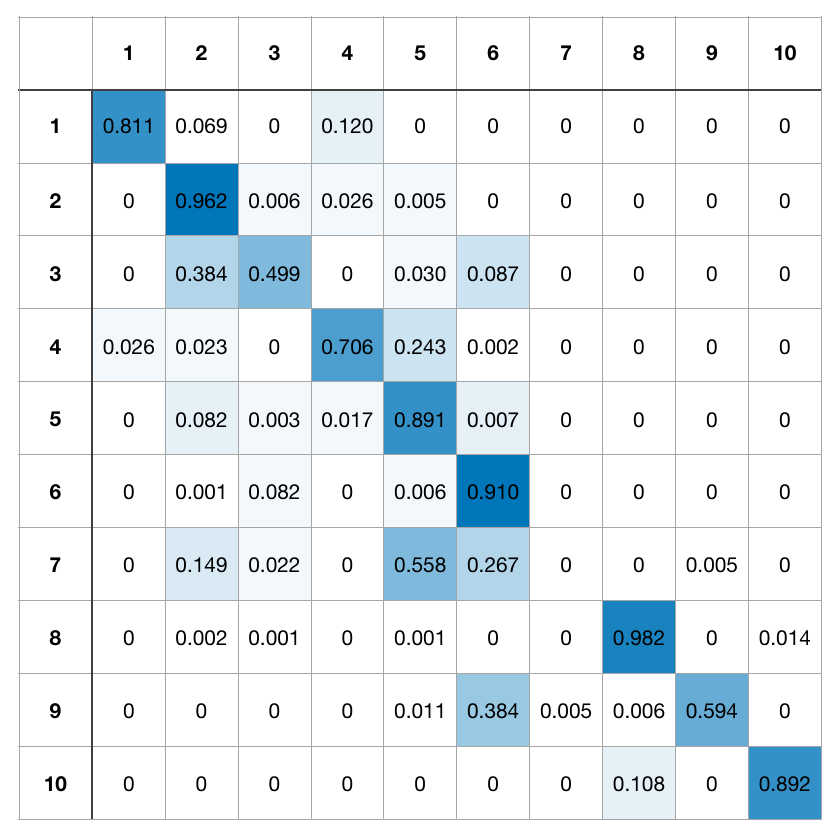}
    \caption{The estimated confusion matrix for the multi-class setting with urban classes is quite similar to the result that we get from all 17 classes.}
    \label{fig:urban_theta}
\end{figure}

Going one step further, we can now look at two separate multi-class classification problems. First, we analyze urban classes, i.e. we only consider images with majority vote for classes {1, ..., 10}, where no confusion with non-urban
classes occurred. Figure \ref{fig:urban_theta} shows the estimated confusion matrix for those images only. We see that performance is similar to the full-class case. For most classes, the probability of correct classification is quite high. Images with only a few votings however could not be detected reliably, see for example classes 1 or 7. 

To look at the experts' voting behaviour, we again conduct the homogeneity analysis for all experts as described in the paper. In this case, a single expert seems to differ from the overall ways of voting but the rest of the voters appear to be more homogeneous.  \\

Table \ref{tab:urban_p} shows the test results for the different cities. On a significant level of 0.05, we can conclude that 10 pairs have different confusion matrices. For the other pairs, we cannot reject the hypothesis of equal voting probabilities. 

\begin{table}
\centering
\small
\begin{tabular}{lll}
\hline
     C1 &      C2 &        p \\
\hline
 berlin & cologne & 0.011 (*) \\
 berlin &  london & 0.017 (*)\\
 berlin &  madrid & 0.794 \\
 berlin &   milan & 0.021 (*)\\
 berlin &  munich & 0.022 (*)\\
 berlin &   paris & 0.928 \\
 berlin &    rome & 0.015 (*)\\
 berlin &  zurich & 0.765 \\
cologne &  london & 0.790 \\
cologne &  madrid & 0.106 \\
cologne &   milan & 0.972 \\
cologne &  munich & 0.886 \\
cologne &   paris & 0.000 (*)\\
cologne &    rome & 0.909\\
cologne &  zurich & 0.079 \\
 london &  madrid & 0.452 \\
 london &   milan & 0.946 \\
 london &  munich & 0.974 \\
 london &   paris & 0.484 \\
 london &    rome & 0.990 \\
 london &  zurich & 0.024 (*)\\
 madrid &   milan & 0.097 \\
 madrid &  munich & 0.053 \\
 madrid &   paris & 0.034 (*)\\
 madrid &    rome & 0.269 \\
 madrid &  zurich & 0.589 \\
  milan &  munich & 0.563 \\
  milan &   paris & 0.353 \\
  milan &    rome & 0.890 \\
  milan &  zurich & 0.724 \\
 munich &   paris & 0.421 \\
 munich &    rome & 1.000 \\
 munich &  zurich & 0.044 (*)\\
  paris &    rome & 0.024 (*)\\
  paris &  zurich & 0.125 \\
   rome &  zurich & 0.112 \\
\hline
\end{tabular}
    \caption{P-values for urban classes.}
    \label{tab:urban_p}
\end{table}

\subsection{Multiclass: nonurban classes only}
\begin{figure}
    \centering
    \includegraphics[width=0.4\textwidth]{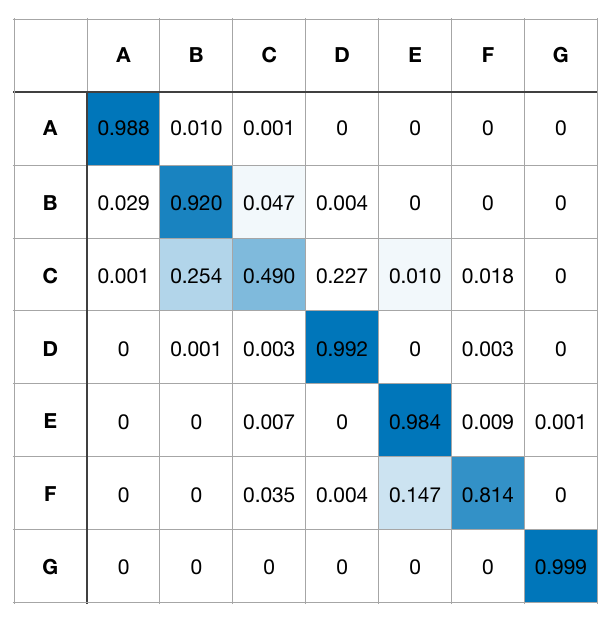}
    \caption{The estimated confusion matrix for the multi-class setting with nonurban classes shows that there are only problems classifying images if we do not have enough data from the respective class.}
    \label{fig:nonurban_theta}
\end{figure}

Second, the procedure is repeated for non-urban labels, i.e.\ only including images with the majority voting for classes {A,...,G}. As Figure \ref{fig:nonurban_theta} shows, we again have difficulties classifying images from classes that did not receive enough votes (like e.g. classes C and F). 

In this setting, the homogeneity analysis for the different experts results in a very similar picture as in the urban case. Again, one expert is not really comparable to the rest of the voters. \\
Looking at the cities separately, the test results show that there are significant differences for 21 pairs (on a significance level of 0.05). 

\begin{table}
    \centering
    \begin{tabular}{lll}
\hline
     C1 &      C2 &     p \\
\hline
 berlin & cologne & 1.000 \\
 berlin &  london & 0.000 (*) \\
 berlin &  madrid & 0.075 \\
 berlin &   milan & 0.226 \\
 berlin &  munich & 0.043 (*)\\
 berlin &   paris & 0.267 \\
 berlin &    rome & 0.007 (*)\\
 berlin &  zurich & 0.000 (*)\\
cologne &  london & 0.001 (*)\\
cologne &  madrid & 0.546 \\
cologne &   milan & 0.204 \\
cologne &  munich & 0.024 (*)\\
cologne &   paris & 0.645 \\
cologne &    rome & 0.001 (*)\\
cologne &  zurich & 0.000 (*)\\
 london &  madrid & 0.002 (*)\\
 london &   milan & 0.044 (*)\\
 london &  munich & 0.262 \\
 london &   paris & 0.006 (*)\\
 london &    rome & 0.125 \\
 london &  zurich & 0.086 \\
 madrid &   milan & 0.469 \\
 madrid &  munich & 0.002 (*)\\
 madrid &   paris & 0.167 \\
 madrid &    rome & 0.002 (*)\\
 madrid &  zurich & 0.002 (*)\\
  milan &  munich & 0.082 \\
  milan &   paris & 0.170 \\
  milan &    rome & 0.019 (*)\\
  milan &  zurich & 0.000 (*)\\
 munich &   paris & 0.009 (*)\\
 munich &    rome & 0.319 \\
 munich &  zurich & 0.001 (*)\\
  paris &    rome & 0.025 (*)\\
  paris &  zurich & 0.000 (*)\\
   rome &  zurich & 0.009 (*)\\
\hline
\end{tabular}

    \caption{P-values for nonurban classes.}
    \label{tab:nonurban_p}
\end{table}